\documentclass[twocolumn,showpacs,preprintnumbers,amsmath,amssymb,superscriptaddress,nofootinbib]{revtex4}
\usepackage{graphicx}
\usepackage{dcolumn}
\usepackage{bm}
\usepackage[chatter]{rotating}
\def\bea{\begin{eqnarray}}
\def\eea{\end{eqnarray}}

\begin{document}


\title{Minimizing statistical and systematic bias in transverse momentum correlations for relativistic heavy-ion collisions}

\affiliation{Department of Physics, The University of Texas at Austin, Austin, Texas 78712 USA}
\author{R. L. Ray and P. Bhattarai}\affiliation{Department of Physics, The University of Texas at Austin, Austin, Texas 78712 USA}

\date{\today}

\begin{abstract}
Two-particle correlation measurements and analysis are an important component of the relativistic heavy-ion physics program. In particular, particle pair-number correlations on two-dimensional transverse momentum ($p_t$) allow unique access to soft, semi-hard and hard-scattering processes in these collisions. Precise measurements of this type of correlation are essential for understanding the dynamics in heavy-ion collisions. However, transverse momentum correlation measurements are especially vulnerable to statistical and systematic biases. In this paper the origins of these large bias effects are explained and mathematical correlation forms are derived from mean-$p_t$ fluctuation quantities in the literature. Monte Carlo simulations are then used to determine the conditions, e.g. multiplicity and collision centrality bin widths, where each correlation form is minimally biased. The ranges of applicability for each correlation quantity are compared. Several are found to reproduce the assumed input correlations with reasonable fidelity over a wide range of conditions encountered in practical analysis of data.
\end{abstract}


\maketitle

\section{Introduction}
\label{SecI}

High-energy collisions between hadron and/or atomic nuclei produce multi-particle final-states for which single- and two-particle number distributions have been measured~\cite{STARspectra,PHENIXspectra,ALICEspectra}. Two-particle correlations, constructed from these distributions, have been shown to be sensitive to the underlying dynamics in the collision process. Parton fragmentation into jets~\cite{Tomjetfrag}, hadronization from soft processes in quantum chromodynamics (QCD)~\cite{LUND}, identical particle quantum interference~\cite{HBT}, parton collectivity (flow)~\cite{flow}, parton-parton quantum interference~\cite{Levin,Pomerons}, resonance decays, and conservation law effects are among the many dynamical processes predicted to contribute to two-particle correlations. The majority of two-particle correlation measurements reported for relativistic heavy-ion collisions at the Relativistic Heavy-Ion Collider (RHIC) and at the Large Hadron Collider (LHC) are angular correlations on sub-spaces $(\phi_1,\phi_2)$, $(\eta_1,\eta_2)$, $(\phi_1 - \phi_2)$, $(\eta_1 - \eta_2)$, or $(\eta_1 - \eta_2,\phi_1 - \phi_2)$, where $\phi$ and $\eta$ are the azimuth and pseudorapidity~\footnote{Pseudorapidity is defined as $\eta = -\ln[\tan(\theta/2)]$, where $\theta$ is the polar scattering angle relative to the beam direction.} of arbitrary particles 1 and 2. Particles are selected within fixed $p_t$ ranges depending on the physics goals of the analysis.

In this paper the complementary correlations on transverse momentum $(p_{t1},p_{t2})$ for a fixed $(\eta,\phi)$ binning scale (bin size) and acceptance range~\cite{ptscale,STARscale} are considered. Measurements of two-particle correlations on $(p_{t1},p_{t2})$ have been reported by experiments NA49~\cite{NA49ptpt,JeffReid}, CERES~\cite{CERESptpt}, and STAR~\cite{JeffReid,JeffQM01,Ayamtmt}. In general, this type of correlation depends on the angular $(\eta,\phi)$ bin scale, acceptance range, and location in $(\eta,\phi)$ space as discussed in Refs.~\cite{STARscale,TrainorMeth,Trainormeanpt}. Here, the $(\eta,\phi)$ bin-scale is fixed at $\Delta\phi = 2\pi$, $\Delta\eta = 2$ for $|\eta| \leq 1$ corresponding to the STAR experiment's time projection chamber (TPC) tracking detector acceptance~\cite{STARNIM,STARTPC} at the RHIC.

For symmetric collision systems, e.g. Au+Au and Pb+Pb, near mid-rapidity the correlations are approximately constant with respect to coordinate $(\eta_1 + \eta_2)$~\cite{AyaCD}. For unpolarized ion + ion collisions the correlations are invariant on coordinate $(\phi_1 + \phi_2)$. Two-particle correlations for these conditions can therefore be considered four-dimensional (4D) functions of variables $p_{t1}$, $p_{t2}$, $\eta_1 - \eta_2$ and $\phi_1 - \phi_2$. Two-dimensional measurements on $(\eta_1 - \eta_2,\phi_1 - \phi_2)$ as a function of 2D $(p_{t1},p_{t2})$, in principle, contain all of the two-particle correlation information. However, as discussed in this paper, angular correlations include an undetermined constant offset~\cite{axialCI} and are therefore incomplete. We will use the relation between two-particle correlations on 2D $(p_{t1},p_{t2})$ space and non-statistical fluctuations in mean-$p_t$~\cite{Trainormeanpt,meanptpaper} to determine the overall, absolute magnitude of the correlations on $(p_{t1},p_{t2})$ and thus resolve the above indeterminacy and allow access to all the information in 4D two-particle correlations.

Two-particle correlations on transverse momentum may provide access to dynamical processes beyond that which can be studied with angular correlations alone. For example, in thermodynamic models, event-wise fluctuations in the final-state temperature of the observed collision system produce fluctuating slopes of the event-wise single-particle $p_t$ distributions resulting in a saddle-shape in the $(p_{t1},p_{t2})$ correlation~\cite{Ayamtmt}. Fluctuating slopes would not produce angular correlations unless they originate in regions with a characteristic angular scale. Another example is the fragmentation of minimum-bias jets~\cite{axialCI} which occurs within a characteristic angular scale and within a relatively local $p_t$ range at intermediate momentum of order 1-4~GeV/$c$~\cite{Tomjetfrag}. Fluctuations in the number of minimum-bias jets and/or the number of charged hadrons per jet cause the intermediate $p_t$ distribution to fluctuate resulting in positive correlations in $(p_{t1},p_{t2})$ along the $p_{t1}=p_{t2}$ diagonal. Angular correlations from minimum-bias jets~\cite{axialCI} determine only the average number of correlated particle-pairs from these processes. Together, angular and $(p_{t1},p_{t2})$ correlations provide access to independent information (i.e. averages and variances) about the event-wise number of correlated particle-pairs from dynamical processes such as jet fragmentation which tend to be localized in both angular and transverse momentum spaces.

In the RHIC and the LHC experiments correlations are measured as functions of global properties of the collision events. Typically events are grouped according to an extensive, event-wise observable such as total charged-particle multiplicity, number of neutrons at zero-degree scattering angle, total transverse energy, etc. which serve as proxies for the degree of overlap, or centrality, between the colliding heavy-ions. In order to achieve sufficient statistical accuracy events must be collected into finite width bins (e.g. centrality bins) in the extensive observable within which the number of produced particles in the collision, or multiplicity, as well as the shape of the single-particle distribution vary.  These variations over finite width centrality bins can {\em bias}, or distort the measurements. This bias is inconsequential for angular correlations~\footnote{Refering to Eq.~(\ref{Eq0}), statistical bias mainly adds a constant offset ($\xi$) to angular correlations and does not affect the analysis of correlation structure. Systematic bias caused by changes in the shape of the single-particle distribution or acceptance within a centrality bin, e.g. dependence of the raw pseudorapidity distribution on collision vertex position within the fiducial volume of the detector, can be minimized by requiring event-mixing within narrow sub-bins.}~\cite{axialCI} but can be quite severe for $(p_{t1},p_{t2})$ correlations~\cite{LizThesis}, being comparable to or larger than the intrinsic correlation structures of interest. The purposes of this work are to derive candidate $(p_{t1},p_{t2})$ correlation measures from non-statistical mean-$p_t$ fluctuation quantities in the literature, estimate the severity of the measurement bias for each form, and determine the range of centrality, multiplicity and kinematics where each correlation quantity is minimally biased.

In the present context {\em bias} refers to any effect which causes the measured correlation to be non-zero when the true correlations vanish, or which distorts the correlation measurement in the presence of true correlations. For example, consider a typical correlation measurement where particle pairs from the same event (sibling pairs) are binned on variable $x$ (e.g. the above coordinate variables) in histogram $N^{\rm sib}(x)$ and mixed-event pairs (the two particles are taken from different events) are collected in $N^{\rm mix}(x)$. Both histograms are averaged over the events. Event-averaged particle pair densities are given by the ratio of the histograms to bin area. Correlation quantity $r(x)-1$~\cite{axialCI} is given by
\bea
r(x)-1 & = & \frac{N^{\rm sib}(x) - N^{\rm mix}(x)}{N^{\rm mix}(x)}
\nonumber \\
 & = & \frac{\overline{N(N-1)} \hat{N}^{\rm sib}(x) - \bar{N}^2
\hat{N}^{\rm mix}(x)}{\bar{N}^2\hat{N}^{\rm mix}(x)}
\nonumber \\
 & = & \frac{\left( \overline{N(N-1)}/\bar{N}^2 \right) \hat{N}^{\rm sib}(x)
- \hat{N}^{\rm mix}(x)}{\hat{N}^{\rm mix}(x)}
\nonumber \\
 & = & (1+\xi) \left( \frac{\hat{N}^{\rm sib}(x) - \hat{N}^{\rm mix}(x)}
{\hat{N}^{\rm mix}(x)} \right) + \xi ,
\label{Eq0}
\eea
where $\bar{N}$ is the mean multiplicity, $\overline{N(N-1)}$ is the mean number of sibling pairs, ``hat'' symbols denote unit-normalized histograms, e.g. $\sum_x \hat{N}^{\rm sib}(x) = 1$, and $\xi \equiv \overline{N(N-1)}/\bar{N}^2 - 1$ where $|\xi| <\!< 1$ if $\bar{N} >\!> 1$ and the range of event-wise multiplicities is $\ll \bar{N}$. The algebraic steps used in going from the first to second line in Eq.~(\ref{Eq0}) are explained in the next section [see Eq.~(\ref{Eq7})] where, for this example, it is assumed that the shapes of the densities do not vary with event multiplicity. Variable $\xi$ is non-zero due to pair counting statistics, a {\em statistical} bias, where factor $(1+\xi)$ is a {\em multiplicative} bias and constant factor $\xi$ on the right-hand side (RHS) of the last line of Eq.~(\ref{Eq0}) is an {\em additive} bias. Variations in the shape of the single-particle distributions within the centrality bin will also cause the numerator in Eq.~(\ref{Eq0}) to not vanish in the absence of true correlations, a {\em systematic} bias. The advantage of $(p_{t1},p_{t2})$ correlations for constraining the 4D correlation measurements is negated by the additive bias $\xi$. For $(p_{t1},p_{t2})$ correlations reported as the number of correlated pairs per final-state particle~\cite{axialCI} the additive bias introduces large, shape distortions in the correlation structures as shown below. A major goal of this paper is to derive $(p_{t1},p_{t2})$ correlation measures for which additive bias effects are negligible.

This paper is organized as follows. In Sec.~\ref{SecII} charge-identified (CID) $(p_{t1},p_{t2})$ correlation quantities are derived from both simple definitions and from mean-$p_t$ fluctuation quantities from the literature. In Sec.~\ref{SecIII} analytic leading-order bias contributions are derived which are due to systematic variations in single-particle distributions and correlation shapes within the finite-width centrality bin. In Sec.~\ref{SecIV} a Monte Carlo simulation model is described which was used to estimate bias effects corresponding to realistic analysis of RHIC data. Simulation results are presented and discussed in Sec.~\ref{SecV}. A summary and conclusions are given in Sec.~\ref{SecVI}.

\section{Transverse momentum correlations}
\label{SecII}

Correlation quantities based on conventional definitions used in angular correlation analysis are discussed first. In Refs.~\cite{Ayamtmt,CLTTom} it was shown that a measure of nonstatistical, event-wise fluctuations in mean-$p_t$ is proportional to the $p_t$-weighted integral of a two-particle, transverse momentum correlation. This is an example of the general relationship between correlations and nonstatistical fluctuations~\cite{ptscale,Trainormeanpt,CLTTom}. Multiple definitions of mean-$p_t$ fluctuation measures can be found in the literature~\cite{meanptpaper,phipt,Voloshin,Fpt,ceres,NA49}. Those which are advocated by experimental collaborations at the CERN Super Proton Synchrotron (SPS), the RHIC, and the LHC are considered here where in each case the corresponding two-particle, transverse momentum correlations for like-sign (LS) and unlike-sign (US) charged-particle pairs are derived.

\subsection{Simple definitions}
\label{SecIIA}

Many of the definitions of correlations in the literature~\cite{AyaCD,axialCI,AyaCI} arbitrarily assume total pair normalization where correlation quantity $(r_{\rm pair}-1)$ on binned space $x$ is given by
\bea
r_{\rm pair}(x)-1 & = & \frac{N^{\rm mix}}{N^{\rm sib}} \frac{h^{\rm sib}(x)}{h^{\rm mix}(x)} -1 \nonumber \\
 & = & \frac{N^{\rm mix}}{N^{\rm sib}} \frac{\left( h^{\rm uncorr}(x) + h^{\rm corr}(x) \right)}{h^{\rm mix}(x)} - 1 \label{Eq1} \\
 & = & \frac{N^{\rm mix}}{N^{\rm sib}} \frac{h^{\rm corr}(x)}{h^{\rm mix}(x)}
 + \left[  \frac{N^{\rm mix}}{N^{\rm sib}} \frac{h^{\rm uncorr}(x)}{h^{\rm mix}(x)} - 1 \right].
\nonumber \\
\label{Eq2}
\eea
In these equations $h^{\rm sib}(x)$ and $h^{\rm mix}(x)$ are histograms of sibling and mixed-event particle pairs, respectively, in bin $x$ ($x$ may represent bins in 1D or 2D angular or $p_t$ sub-spaces), $h^{\rm sib}(x) = h^{\rm uncorr}(x) + h^{\rm corr}(x)$ corresponding to the number of uncorrelated (random) and correlated pairs, and the total pair counts are given by $N^{\rm sib} = \sum_x h^{\rm sib}(x)$ and $N^{\rm mix} = \sum_x h^{\rm mix}(x)$.

The conventional definition of the unit-normalized two-particle density~\cite{Feshbach,Cramer}, $\hat{\rho}(x_1,x_2)$, is given by
\bea
\hat{\rho}(x_1,x_2) & = & \hat{\rho}(x_1) \hat{\rho}(x_2) + C(x_1,x_2)
\label{Eq3}
\eea
where the one-particle density, $\hat{\rho}(x_1)$, is the marginal distribution of the two-particle density, $\hat{\rho}(x_1) = \int dx_2 \hat{\rho}(x_1,x_2)$, and $C(x_1,x_2)$ is the two-particle correlation density. From this definition we find that
\bea
\int dx_2 C(x_1,x_2) & = & \int dx_1 C(x_1,x_2) = 0.
\label{Eq4}
\eea
However, the integral of $C(x_1,x_2)$ over a reduced portion of the full space (e.g. the detector acceptance $\Delta x$), given by
\bea
\int_{\Delta x} dx_2 C(x_1,x_2) & \neq & 0,
\label{Eq5}
\eea
does not vanish in general. For the bin counts in Eqs.~(\ref{Eq1}) and (\ref{Eq2}) the preceding nonvanishing integral requires that $\sum_x h^{\rm corr}(x) \neq 0$ and therefore $\sum_x h^{\rm uncorr}(x) \neq N^{\rm sib}$. It follows that the factor in square brackets in Eq.~(\ref{Eq2}) does not vanish in general and is approximately a constant $\lambda$ over the domain of $x$, where $|\lambda| <\!< 1$ if $h^{\rm corr}(x) <\!< h^{\rm uncorr}(x)$. Both quantities on the RHS of Eq.~(\ref{Eq2}) are small $(<\!<1)$, but may be comparable, i.e. the arbitrary $\lambda$ could be of the order of the correlation amplitude.

It is conventional~\cite{Trainormeanpt,axialCI} to report angular correlations as a normalized covariance (Pearson's normalized covariance~\cite{Pearson}) by multiplying $[r_{\rm pair}(x)-1]$ in Eq.~(\ref{Eq1}) by a single particle quantity or histogram, $\sqrt{\rho_{\rm ref}(x)}$, where
\bea 
\sqrt{\rho_{\rm ref}(x)} \left[ r_{\rm pair}(x) - 1 \right] & = & \sqrt{\rho_{\rm ref}(x)} \frac{N^{\rm mix}}{N^{\rm sib}} \frac{h^{\rm corr}(x)}{h^{\rm mix}(x)}
\nonumber \\
  & + &  \lambda \sqrt{\rho_{\rm ref}(x)}.
\label{Eq6}
\eea
For angular correlations from symmetric collision systems (e.g. p+p, Au+Au, Pb+Pb) at mid-rapidity the {\em prefactor}, $\sqrt{\rho_{\rm ref}(x)}$ (see also Sec.~\ref{SecIIG}), is approximately constant and is given by $d^2N_{\rm ch}/d\eta d\phi$~\cite{axialCI} where $N_{\rm ch}$ is the charged-particle multiplicity. Factor $\lambda \sqrt{\rho_{\rm ref}(x)}$ contributes an unknown, constant offset to the angular correlations meaning that only the non-constant angular correlation structures are physically significant as explained in Ref.~\cite{axialCI}.

For transverse momentum correlations the prefactor is given by $\sqrt{(d^2N_{\rm ch}/dp_{t1}d\eta_1)(d^2N_{\rm ch}/dp_{t2}d\eta_2)}$ which varies exponentially with $(p_{t1},p_{t2})$. This correlation per final-state particle measure provides much greater visual access to the correlation structures at low and intermediate $p_t$ less than a few GeV/$c$. In this case the structure of the unknown factor $\lambda \sqrt{\rho_{\rm ref}(p_{t1},p_{t2})}$ may be comparable to or larger than the true correlations, making the $(p_{t1},p_{t2})$ pair-normalized correlations unreliable. Equation~(\ref{Eq6}) and bias factor $\lambda$ apply to both LS and US charged particle-pairs.

Another correlation definition, referred to in the Introduction, invokes event averaging where the sibling pair histogram is given by
\begin{widetext}
\bea
N^{\rm sib}(x) & = & \frac{1}{\epsilon} \sum_{j=1}^{\epsilon} n^{\rm sib}_j(x)
 = \sum_m \frac{\epsilon_m}{\epsilon} \frac{1}{\epsilon_m} \sum_{j=1}^{\epsilon_m} n^{\rm sib}_{jm}(x) 
  =  \sum_m \frac{\epsilon_m}{\epsilon} \bar{n}^{\rm sib}_m(x)
\nonumber \\ 
& =  & 
\sum_m \frac{\epsilon_m}{\epsilon} m(m-1) \hat{\bar{n}}^{\rm sib}_m(x)
  \approx \sum_{\delta_m} \frac{\epsilon_m}{\epsilon} ( \bar{N} + \delta_m)(\bar{N} + \delta_m - 1) \hat{\bar{n}}^{\rm sib}(x)
\nonumber \\
 &  \approx & \left[ \bar{N}( \bar{N} - 1) + \sigma^2_N \right]
\hat{\bar{n}}^{\rm sib}(x) 
\label{Eq7}
\eea
where in the first line $\epsilon$ is the number of collision events, $j$ is the event index, $n^{\rm sib}_j(x)$ is the number of sibling pairs in event $j$ in bin $x$, $m$ is an event multiplicity value within the centrality range, $\epsilon_m$ is the number of events which have multiplicity $m$, and $\bar{n}^{\rm sib}_m(x)$ is an average over all events with multiplicity $m$. In the second line event-wise pair count $m(m-1)$ includes both permutations of particles 1 and 2, $\hat{\bar{n}}^{\rm sib}_m(x)$ is normalized to unity where $\hat{\bar{n}}^{\rm sib}_m(x) = \bar{n}^{\rm sib}_m(x)/\sum_x\bar{n}^{\rm sib}_m(x)$, $\bar{N} = \sum_m (\epsilon_m/\epsilon) m$ is the mean multiplicity, $m = \bar{N} + \delta_m$, and in the last line $\sigma^2_N$ is the variance of the multiplicity distribution of the $\epsilon$ events given by $\sum_m (\epsilon_m/\epsilon) (m-\bar{N})^2$. In the second line of Eq.~(\ref{Eq7}) the possible multiplicity dependence of the shape of $\hat{\bar{n}}^s_m(x)$ was neglected.

The mixed-event pair histogram is given by
\bea
N^{\rm mix}(x) & = & \frac{1}{\epsilon_{\rm mix}} \sum_{j \neq j^{\prime}}
\left[ n_j(x_1) n_{j^{\prime}}(x_2) \right]_{(x)}
  =  \bar{N}^2
\left[ \hat{\bar{n}}(x_1) \hat{\bar{n}}(x_2) \right]_{(x)}
\label{Eq8}
\eea
where $\epsilon_{\rm mix}$ is the number of pairs of mixed events used in the summation, $n_j(x_1)$ and $n_{j^{\prime}}(x_2)$ are the binned single-particle counts in events $j$ and $j^{\prime}$ where $j \neq j^{\prime}$, and notation $\left[ \hat{\bar{n}}(x_1) \hat{\bar{n}}(x_2) \right]_{(x)}$ means that all mixed-event particle-pairs which contribute to pair-wise bin $x$ are included in the summation. For transverse momentum correlations this factor is explicitly given by $\hat{\bar{n}}(p_{t1}) \hat{\bar{n}}(p_{t2})$ where in this context $p_{t1}$ and $p_{t2}$ represent $p_t$ bins.

The normalized, event-averaged correlation is given by
\bea
\sqrt{\rho_{\rm ref}}(r_{\rm event}-1) & = & \sqrt{\frac{d^2N_{\rm ch}}{dp_{t1}d\eta_1} 
                     \frac{d^2N_{\rm ch}}{dp_{t2}d\eta_2}}
     \frac{N^{\rm sib}(p_{t1},p_{t2})-N^{\rm mix}(p_{t1},p_{t2})}
          {N^{\rm mix}(p_{t1},p_{t2})}.
\label{Eq9}
\eea
Multiplying $N^{\rm sib}(x)$ by $\bar{N}/(\bar{N}-1)$ removes the trivial pair counting difference between sibling and mixed-event pairs [see Eqs.~(\ref{Eq7}) and (\ref{Eq8})]. Equation~(\ref{Eq9}) can then be re-expressed as
\bea
\sqrt{\rho_{\rm ref}}(r_{\rm event}-1) & = & \sqrt{\frac{d^2N_{\rm ch}}{dp_{t1}d\eta_1}
                     \frac{d^2N_{\rm ch}}{dp_{t2}d\eta_2}}
     \frac{\frac{\bar{N}}{\bar{N} - 1} N^{\rm sib}(p_{t1},p_{t2})-N^{\rm mix}(p_{t1},p_{t2})}
          {N^{\rm mix}(p_{t1},p_{t2})}
\nonumber \\
 & \approx & \sqrt{\frac{d^2N_{\rm ch}}{dp_{t1}d\eta_1}
                     \frac{d^2N_{\rm ch}}{dp_{t2}d\eta_2}}
 \left[ \frac{\hat{\bar{n}}^{\rm sib}(p_{t1},p_{t2}) - \hat{\bar{n}}(p_{t1}) \hat{\bar{n}} (p_{t2})}{\hat{\bar{n}}(p_{t1}) \hat{\bar{n}} (p_{t2})}
 + \frac{\sigma^2_N}{\bar{N}(\bar{N} - 1)}
  \frac{\hat{\bar{n}}^{\rm sib}(p_{t1},p_{t2})}
       {\hat{\bar{n}}(p_{t1}) \hat{\bar{n}} (p_{t2})} \right] .
\label{Eq10}
\eea
\end{widetext}
In the absence of true correlations $\hat{\bar{n}}^{\rm sib}(p_{t1},p_{t2})$ equals $\hat{\bar{n}}(p_{t1}) \hat{\bar{n}}(p_{t2})$. However, $(r_{\rm event}-1)$ is not zero in that limit due to the additive bias term proportional to $\sigma^2_N$ which is determined by the multiplicity or centrality bin width. The bias is approximately $\sqrt{\rho_{\rm ref}} (\sigma_N/\bar{N})^2$ and can be much larger than the typical correlations as shown in Sec.~\ref{SecV} where for heavy-ion collisions $|\hat{\bar{n}}^{\rm sib}(p_{t1},p_{t2}) / (\hat{\bar{n}}(p_{t1}) \hat{\bar{n}}(p_{t2})) - 1|$ is of order $10^{-3}$ to $10^{-2}$~\cite{Ayamtmt}. Alternatively, the ratio $r = N^{\rm sib}(x)/N^{\rm mix}(x)$ could be normalized by the factor $\bar{N}^2/[\bar{N}(\bar{N} - 1) + \sigma^2_N]$ which produces the same result and possible distortion as found for the above pair-normalization method. Equation~(\ref{Eq10}) directly applies to non-charge-identified particle-pairs and to LS pairs. For US pairs statistical bias persists where $\sigma^2_N/[\bar{N}(\bar{N} - 1)]$ in Eq.~(\ref{Eq10}) is replaced with $cov[(n^+ - \bar{N}^+)(n^- - \bar{N}^-)]/(\bar{N}^+ \bar{N}^-)$, the normalized covariance between positive and negative charged-particle number fluctuations.

\subsection{Correlation derived from $\Delta \sigma^2_{p_t:m}$}
\label{SecIIB}

Several authors have proposed mean-$p_t$ fluctuation quantities which minimize statistical bias, all of which rely on the scale invariance (i.e. angular bin-size invariance), in the absence of correlations, of the quantity $m\sigma^2_{\langle p_t \rangle}$, where $m$ and $\sigma^2_{\langle p_t \rangle}$ are respectively the multiplicity and variance (defined below) of mean-$p_t$ fluctuations within the angular bin. This scale invariance is a consequence of the central limit theorem (CLT)~\cite{CLTTom,CLT}. Non-statistical fluctuations, which correspond to correlations, break this scale invariance causing the difference $[(m\sigma^2_{\langle p_t \rangle})_{\delta x_2} - (m\sigma^2_{\langle p_t \rangle})_{\delta x_1}]$ to be non-zero where subscripts $\delta x_1$ and $\delta x_2$ refer to different angular bin-sizes, or scales. However, there is not a unique method for implementing this scale difference quantity in the definitions of non-statistical mean-$p_t$ fluctuation measures. For example, difference $[(m\sigma^2_{\langle p_t \rangle})_{\delta x_2} - (m\sigma^2_{\langle p_t \rangle})_{\delta x_1}]$ can be multiplied by arbitrary powers of $m$ in order to minimize bias due to the $m$-dependence in the non-statistical fluctuations. A linear width-difference, $\sqrt{(m\sigma^2_{\langle p_t \rangle})_{\delta x_2}} - \sqrt{(m\sigma^2_{\langle p_t \rangle})_{\delta x_1}}$, could also be used. This ambiguity allows multiple forms for mean-$p_t$ fluctuation quantities to be defined.

In Refs.~\cite{JeffReid,LiuTrainor} Liu, Trainor and Reid proposed the quantity $\Delta \sigma^2_{p_t:m}$ based directly on the above variance difference. This quantity was used by the STAR Collaboration in the analysis of Au+Au collisions at $\sqrt{s_{\rm NN}}$ = 130~GeV~\cite{meanptpaper}. Subscript $p_t:m$ emphasizes that this quantity measures non-statistical fluctuations of transverse momentum with negligible contribution from fluctuations in multiplicity ($m$)~\cite{meanptpaper}. This quantity was designed to eliminate bias when $[(m\sigma^2_{\langle p_t \rangle})_{\delta x_2} - (m\sigma^2_{\langle p_t \rangle})_{\delta x_1}]$ varies as $(f_0 + mf_1)$ in the presence of non-statistical fluctuations. For non-charge-identified particles this quantity at the acceptance scale is given by
\bea
\Delta \sigma^2_{p_t:m} & = & \frac{1}{\epsilon} \sum_{j=1}^{\epsilon} n_j
\left( \langle p_t \rangle_j - \hat{p}_t \right)^2
- \sigma^2_{\hat{p}_t}
\label{Eq11}
\eea
where $n_j$ is the multiplicity within the acceptance for event $j$, and event-wise mean-$p_t$, inclusive mean-$p_t$ and inclusive $p_t$ variance are respectively given by
\bea
\langle p_t \rangle_j & = & \frac{1}{n_j} \sum_{i=1}^{n_j} p_{t,ji},
\label{Eq12} \\
\hat{p}_t & = & \frac{1}{\epsilon \bar{N}} \sum_{j=1}^{\epsilon}
\sum_{i=1}^{n_j} p_{t,ji},
\label{Eq13} \\
\sigma^2_{\hat{p}_t} & = & \frac{1}{\epsilon \bar{N}} \sum_{j=1}^{\epsilon}
\sum_{i=1}^{n_j} \left( p_{t,ji} - \hat{p}_t  \right)^2.
\label{Eq14}
\eea
The inclusive $p_t$ variance represents $(m\sigma^2_{\langle p_t \rangle})_{\delta x_1}$ in the limit of very small bin sizes where occupied bins contain exactly one particle and only occupied bins are included in the summations~\cite{JeffReid,meanptpaper,LiuTrainor}.
In this paper angle brackets ``$ \langle$\,$\rangle$'' denote event-wise averages and over-lines denote averages over an event collection.

In the absence of non-statistical fluctuations $\Delta \sigma^2_{p_t:m}$ equals zero, where
\begin{widetext}
\bea
\Delta \sigma^2_{p_t:m} & = & \sum_m \frac{\epsilon_m}{\epsilon} m \frac{1}{\epsilon_m}
\sum_{j=1}^{\epsilon_m} \left( \langle p_t \rangle_j - \hat{p}_t \right)^2
- \sigma^2_{\hat{p}_t}
  =  \sum_m \frac{\epsilon_m}{\epsilon} m \sigma^2_{\langle p_t \rangle} - \sigma^2_{\hat{p}_t} \rightarrow 0
\label{Eq15}
\eea
using the CLT result $m\sigma^2_{\langle p_t \rangle} = \sigma^2_{\hat{p}_t}$ where $\sigma^2_{\langle p_t \rangle}$ is the variance of the distribution of event-wise mean-$p_t$ for events with $m$ particles in the angular bin. The expression for $\Delta \sigma^2_{p_t:m}$ in Eq.~(\ref{Eq11}) can be expanded in terms of particle pairs by substituting the definitions in Eqs.~(\ref{Eq12}) - (\ref{Eq14}), collecting terms proportional to sums of pairs of particles from the same event (siblings) and sums of pairs of particles from different events (mixed-event pairs), and assuming a large number of events $\epsilon >\!> 1$. The result is given by
\bea
\bar{N}\Delta \sigma^2_{p_t:m} & = & \frac{1}{\epsilon} \sum_{j=1}^{\epsilon}
\frac{\bar{N}}{n_j} \sum_{i \neq i^{\prime} = 1}^{n_j} p_{t,ji} p_{t,ji^{\prime}}
- \frac{\bar{N} - 1}{\bar{N}} \frac{1}{\epsilon_{\rm mix}}
\sum_{j \neq j^{\prime}} \sum_{i=1}^{n_j} \sum_{i^{\prime} = 1}^{n_{j^{\prime}}}
p_{t,ji} p_{t,j^{\prime} i^{\prime}}
  +  \frac{1}{\epsilon} \sum_{j=1}^{\epsilon}
\left( \frac{\bar{N}}{n_j} - 1 \right) \sum_{i=1}^{n_j} p_{t,ji}^2 .
\label{Eq16}
\eea
\end{widetext}
The last term in Eq.~(\ref{Eq16}) is a self-pair term which is non-vanishing when average $p_t^2$ is correlated with event multiplicity, but vanishes otherwise. It may contribute to the fluctuation measure but does not contribute to the correlation.

The particle sums, when binned on 2D transverse momentum, can be expressed as
\bea
\sum_{i \neq i^{\prime} = 1}^{n_j} p_{t,ji} p_{t,ji^{\prime}} & = &
\sum_{k,l} p_{t,k} p_{t,l} n_{j,kl}^{\rm sib}
\label{Eq17} \\
 \sum_{i=1}^{n_j} \sum_{i^{\prime} = 1}^{n_{j^{\prime}}}
p_{t,ji} p_{t,j^{\prime} i^{\prime}} & = & \sum_{k,l} p_{t,k} p_{t,l}
n_{jk} n_{j^{\prime} l}
\label{Eq18}
\eea
where subscripts $k,l$ are transverse momentum bin indices, $p_{t,k}$ and $p_{t,l}$ are the average $p_t$ within those bins (approximately $p_t$ at the bin centers), $n_{j,kl}^{\rm sib}$ is the number of sibling pairs in 2D bin $(k,l)$ in event $j$, and $n_{jk}$ and $n_{j^{\prime} l}$ are the number of particles in $p_t$ bins $k$ and $l$ in events $j$ and $j^{\prime}$, respectively. By substituting Eqs.~(\ref{Eq17}) and (\ref{Eq18}) into Eq.~(\ref{Eq16}) and omitting the self-pair term, the relationship between the mean-$p_t$ fluctuation measure and the two-particle correlation for this case can be expressed as
\bea
\bar{N} \Delta \sigma^2_{p_t:m} & \approx & \sum_{k,l} p_{t,k} p_{t,l}
\Delta N_{kl,\Delta \sigma^2}
\label{Eq19} \\
\Delta N_{kl,\Delta \sigma^2} & = & \frac{1}{\epsilon} \sum_{j=1}^{\epsilon}
\frac{\bar{N}}{n_j} n_{j,kl}^{\rm sib} -  \frac{\bar{N} - 1}{\bar{N}} \frac{1}{\epsilon_{\rm mix}}
\sum_{j \neq j^{\prime}} n_{jk} n_{j^{\prime} l} .
\nonumber \\
\label{Eq20}
\eea
For like-sign charged-particle pairs ($++$ and $--$) the preceding equation can immediately be written as
\bea
\Delta N_{kl,\Delta \sigma^2}^{\pm \pm} & = &
\frac{1}{\epsilon} \sum_{j=1}^{\epsilon} \frac{\bar{N}^{\pm}}{n_j^{\pm}}
n_{j,kl}^{{\rm sib}\pm \pm}
\nonumber \\
 &  - &  \frac{\bar{N}^{\pm} - 1}{\bar{N}^{\pm}} \frac{1}{\epsilon_{\rm mix}}
\sum_{j \neq j^{\prime}} n_{jk}^{\pm} n_{j^{\prime} l}^{\pm}.
\label{Eq21}
\eea

From Ref.~\cite{meanptpaper} the mean-$p_t$ fluctuation measure for unlike-sign charged-particle pairs is
\bea
\Delta \sigma^{2,US}_{p_t:m} & = & \frac{1}{\epsilon}  \sum_{j=1}^{\epsilon}
\sqrt{n_j^+ n_j^-} \left( \langle p_t^{\pm} \rangle_j - \hat{p}_t^{\pm} \right)
\left( \langle p_t^{\mp} \rangle_j - \hat{p}_t^{\mp} \right) .
\nonumber \\
\label{Eq22}
\eea
After multiplying by $\sqrt{\bar{N}^+ \bar{N}^-}$ and using the CID versions of the summations in Eqs.~(\ref{Eq12}), (\ref{Eq13}), (\ref{Eq17}) and (\ref{Eq18}), the unlike-sign charged-particle pair correlation can be expressed as
\begin{widetext}
\bea
\Delta N_{kl,\Delta \sigma^2}^{\pm \mp} & = &
\frac{1}{\epsilon} \sum_{j=1}^{\epsilon}
\sqrt{\frac{\bar{N}^+ \bar{N}^-}{n_j^+ n_j^-}} 
n_{j,kl}^{{\rm sib}\pm \mp}
  -  \frac{1}{\epsilon_{\rm mix}} \sum_{j^{\prime} \neq j^{\prime \prime}}
\left[ \sqrt{ \frac{\bar{N}^{\pm} n_{j^{\prime}}^{\mp}}{\bar{N}^{\mp} n_{j^{\prime}}^{\pm}}}
+ \sqrt{ \frac{\bar{N}^{\mp} n_{j^{\prime \prime}}^{\pm}}{\bar{N}^{\pm} n_{j^{\prime \prime}}^{\mp}}}
- \overline{ \sqrt{ \frac{ n_j^+ n_j^-}{\bar{N}^+ \bar{N}^-} } }
\right] n_{j^{\prime}k}^{\pm} n_{j^{\prime \prime} l}^{\mp}
\label{Eq23}
\eea
\end{widetext}
where the over-lined quantity in the mixed-event summation is averaged over all events $j = 1,2, \cdots \epsilon$. In obtaining the second weight factor in the mixed-event expression summation indices $j^{\prime}$ and $j^{\prime \prime}$ were interchanged. Including the weight factors in Eqs.~(\ref{Eq21}) and (\ref{Eq23}) is essential for eliminating the finite bin-width statistical bias. Note that all weight factors equal unity when the CID event multiplicities are constant.

The form of Eq.~(\ref{Eq20}) for nonidentified particles suggests the following (simple) CID expression where the sibling-pair term and the single-particle terms with CID labels $a,b$ are written out as
\bea
n_{j,kl}^{\rm sib} & = & \sum_{a = \pm} \sum_{b = \pm} n_{j,kl}^{{\rm sib},ab}
\label{Eq24} \\
n_{jk} & = & \sum_{a = \pm} n_{jk}^a .
\label{Eq25}
\eea
The resulting, alternate CID form for the correlation is given by
\bea
\Delta N_{kl,{\rm alt}}^{ab} & = & \frac{1}{\epsilon} \sum_{j=1}^{\epsilon}
\frac{\bar{N}}{n_j} n_{j,kl}^{{\rm sib},ab}
- \frac{\bar{N} - 1}{\bar{N}}
\frac{1}{\epsilon_{\rm mix}} \sum_{j \neq j^{\prime}} n_{jk}^a n_{j^{\prime}l}^b
\nonumber \\
\label{Eq26}
\eea
for $a,b = \pm,\pm$. In Sec.~\ref{SecV} it will be shown that this correlation definition is strongly biased; only the charge-nonidentified form in Eq.~(\ref{Eq20}) is useful.

The CERES Collaboration introduced a mean-$p_t$ fluctuation quantity $\sigma^2_{p_t,{\rm dyn,Ceres}}$ in Ref.~\cite{ceres} at about the same time $\Delta \sigma^2_{p_t:m}$ was being developed by Liu, Trainor and Reid. It is algebraically identical to $\Delta \sigma^2_{p_t:m}/\bar{N}$ and therefore leads to the same correlation quantities given in Eqs.~(\ref{Eq21}) and (\ref{Eq23}).

\subsection{Correlation derived from $\Phi_{p_t}$}
\label{SecIIC}

A mean-$p_t$ fluctuation width difference quantity $\Phi_{p_t}$~\cite{phipt} is defined as
\bea
\Phi_{p_t} & = & \sqrt{ \overline{Z^2}/\bar{N}} - \sigma_{\hat{p}_t}
\label{Eq27} \\
\overline{Z^2} & = & \frac{1}{\epsilon} \sum_{j=1}^{\epsilon} 
\sum_{i,i^{\prime} = 1}^{n_j} \left( p_{t,ji} - \hat{p}_t \right)
                              \left( p_{t,ji^{\prime}} - \hat{p}_t \right)
\nonumber \\
 & = & \frac{1}{\epsilon} \sum_{j=1}^{\epsilon} n_j^2
       \left( \langle p_t \rangle_j - \hat{p}_t \right)^2.
\label{Eq28}
\eea
Direct conversion of $\Phi_{p_t}$ into a form proportional to a weighted integral of the correlation is complicated by the square-root in  Eq.~(\ref{Eq27}) and the linear form of $\Phi_{p_t}$ which is based on a fluctuation width difference as opposed to a variance difference which was used in the preceding sub-section. An approximate quantity can be defined which depends on variance differences similar to that used for $\Delta \sigma^2_{p_t:m}$. Multiplying Eq.~(\ref{Eq27}) by $\left( \sqrt{ \overline{Z^2}/\bar{N}} + \sigma_{\hat{p}_t} \right)$ yields
\bea
\Phi_{p_t} \left[\sqrt{ \overline{Z^2}/\bar{N}} + \sigma_{\hat{p}_t} \right]
 & = & \overline{Z^2}/\bar{N} - \sigma^2_{\hat{p}_t}
\label{Eq29} 
\eea
and then substituting from Eq.~(\ref{Eq27}) into the factor on the left-hand-side (LHS) results in
\bea
\Phi_{p_t} \left[ \Phi_{p_t} + 2\sigma_{\hat{p}_t} \right]
& \equiv & 2 \sigma_{\hat{p}_t} \Phi_{p_t}^{(0)} ,
\label{Eq30}
\eea
where the RHS defines approximate measure $\Phi_{p_t}^{(0)}$. For heavy-ion collisions $\Phi_{p_t} <\!< \sigma_{\hat{p}_t}$~\cite{phipt} and solving Eq.~(\ref{Eq30}) for $\Phi_{p_t}$ yields the rapidly converging expansion
\bea
\Phi_{p_t} & \approx & \Phi_{p_t}^{(0)} \left[1 - \Phi_{p_t}^{(0)}/(2\sigma_{\hat{p}_t}) + \cdots \right]
\label{Eq31}
\eea
when $\Phi_{p_t}^{(0)} <\!< \sigma_{\hat{p}_t}$. From Eqs.~(\ref{Eq28})-(\ref{Eq30}) we obtain
\bea
\Phi_{p_t}^{(0)} & = & \left[ \overline{Z^2}/\bar{N} - \sigma^2_{\hat{p}_t}
\right] /(2\sigma_{\hat{p}_t})
\nonumber \\
 & = & \frac{1}{2\sigma_{\hat{p}_t} \bar{N} \epsilon} \sum_{j=1}^{\epsilon}
\left[ n_j^2 \left( \langle p_t \rangle_j - \hat{p}_t \right)^2
-n_j \sigma^2_{\hat{p}_t} \right].
\label{Eq32}
\eea
Equation~(\ref{Eq32}) can be directly applied to like-sign pairs $(++)$ and $(--)$. Quantity $\Phi_{p_t}^{(0)}$ includes an additional factor $n_j$ compared to $\Delta \sigma^2_{p_t:m}$ in Eq.~(\ref{Eq11}). The STAR Collaboration adopted the quantity $2\sigma_{\hat{p}_t}\Phi_{p_t}^{(0)}$ for the scale-dependent fluctuation analysis in Refs.~\cite{ptscale,STARscale}.

For unlike-sign charged-particle pairs quantity $\overline{Z^2}$ in Eq.~(\ref{Eq28}) is evaluated for $(\pm \mp)$ pairs where the variance $\sigma_{\hat{p}_t}^2$ in Eq.~(\ref{Eq32}) is not included as was the case for $\Delta \sigma^{2,US}_{p_t:m}$ [see Eq.~(\ref{Eq22})], and the scaling factors $\sigma_{\hat{p}_t}$ and $\bar{N}$ are replaced with geometric means as in Eq.~(\ref{Eq22}). The result from Eqs.~(\ref{Eq28}) and (\ref{Eq32}) is
\bea
\Phi_{p_t}^{(0)+-} & = & \frac{1}{2\sqrt{\sigma^+_{\hat{p}_t} \sigma^-_{\hat{p}_t} \bar{N}^+ \bar{N}^-} }
\nonumber \\
 & \times & \frac{1}{\epsilon} \sum_{j=1}^{\epsilon} n_j^+ n_j^- 
\left( \langle p_t^+ \rangle_j - \hat{p}_t^+ \right)
\left( \langle p_t^- \rangle_j - \hat{p}_t^- \right).
\nonumber \\
\label{Eq33}
\eea

The LS and US correlations are derived by substituting the explicit summations for $\langle p_t^{\pm} \rangle_j$, $\hat{p}_t^{\pm}$ and $(\sigma^{\pm}_{\hat{p}_t})^2$ into Eqs.~(\ref{Eq32}) and (\ref{Eq33}), collecting sibling and mixed-event pair terms, using the $p_t$ binning in Eqs.~(\ref{Eq17}) and (\ref{Eq18}), and factoring out constants. The results are given by
\begin{widetext}
\bea
\Delta N^{\pm \pm}_{kl,\Phi} & =  &
\frac{1}{\epsilon} \sum_{j=1}^{\epsilon}
n_{j,kl}^{{\rm sib}\pm \pm}
  -  \frac{1}{\epsilon_{\rm mix}}
\sum_{j^{\prime} \neq j^{\prime \prime}}
\left[ \frac{ n_{j^{\prime}}^{\pm} - 1}{\bar{N}^{\pm}}
  +    \frac{ n_{j^{\prime \prime}}^{\pm} - 1}{\bar{N}^{\pm}}
  - \frac{\overline{n_j^{\pm} (n_j^{\pm} - 1)}}{\bar{N}^{\pm^2}} \right]
n_{j^{\prime} k}^{\pm} n_{j^{\prime \prime} l}^{\pm}
\label{Eq34} \\
\Delta N^{\pm \mp}_{kl,\Phi} & =  & 
\frac{1}{\epsilon} \sum_{j=1}^{\epsilon}
n_{j,kl}^{{\rm sib}\pm \mp}
  -  \frac{1}{\epsilon_{\rm mix}}
\sum_{j^{\prime} \neq j^{\prime \prime}}
\left[ \frac{ n_{j^{\prime}}^{\mp}}{\bar{N}^{\mp}}
  +    \frac{ n_{j^{\prime \prime}}^{\pm}}{\bar{N}^{\pm}}
  - \frac{\overline{n_j^+n_j^-}}{\bar{N}^+ \bar{N}^-} \right]
n_{j^{\prime} k}^{\pm} n_{j^{\prime \prime} l}^{\mp}
\label{Eq35}
\eea
where self-pair terms cancel in this case.

The ALICE Collaboration defined a mean-$p_t$ fluctuation quantity $C_{p_t}$\cite{ALICECpt} given by
\bea
C_{p_t} & = & \frac{1}{N_{\rm pairs}} \frac{1}{\epsilon} \sum_{j=1}^{\epsilon}
\sum_{i \neq i^{\prime} = 1}^{n_j} \left( p_{t,ji} - \hat{p}_t \right)
                              \left( p_{t,ji^{\prime}} - \hat{p}_t \right),
\label{Eq36}
\eea
a variance difference as shown in Ref.~\cite{Trainormeanpt}, where $N_{\rm pairs}$ is the event-average number of particle pairs. The correlations are derived by inserting the expansions for $\hat{p}_t$ and collecting sibling and mixed-event pair terms for like-sign and unlike-sign charged-particle pairs as above. The resulting correlations are the same as those derived for $\Phi^{(0)}_{p_t}$ in Eqs.~(\ref{Eq34}) and (\ref{Eq35}).

\subsection{Correlation derived from $\sigma^2_{p_t,{\rm dynamical}}$}
\label{SecIID}

Mean-$p_t$ fluctuation quantity $\sigma^2_{p_t,{\rm dynamical}}$~\cite{Voloshin} is defined for like-sign and unlike-sign particle pairs, using a variance difference form (see Ref.~\cite{Trainormeanpt}) given by
\bea
\sigma^{2\pm \pm }_{p_t,{\rm dynamical}} & = & 
\frac{1}{\epsilon} \sum_{j=1}^{\epsilon} \frac{1}{n_j^{\pm}(n_j^{\pm} - 1)}
  \sum_{i \neq i^{\prime} = 1}^{n_j^{\pm}} \left( p_{t,ji}^{\pm} - \hat{p}_t^{\pm} \right)
                   \left( p_{t,ji^{\prime}}^{\pm} - \hat{p}_t^{\pm} \right)
\label{Eq37} \\
\sigma^{2\pm \mp }_{p_t,{\rm dynamical}} & = &
\frac{1}{\epsilon} \sum_{j=1}^{\epsilon} \frac{1}{n_j^+ n_j^-}
\sum_{i=1}^{n_j^{\pm}} \sum_{i^{\prime} = 1}^{n_j^{\mp}}
\left( p_{t,ji}^{\pm} - \hat{p}_t^{\pm} \right)
\left( p_{t,ji^{\prime}}^{\mp} - \hat{p}_t^{\mp} \right).
\label{Eq38}
\eea
This quantity is directly proportional to a weighted integral of a two-particle correlation. Following the same steps as in Sec.~\ref{SecIIB} the corresponding correlations are given by
\bea
\Delta N^{\pm \pm}_{kl,\sigma-{\rm dyn}} & = & 
\frac{1}{\epsilon} \sum_{j=1}^{\epsilon}
\frac{\bar{N}^{\pm^2}}{n_j^{\pm}(n_j^{\pm} - 1)}
     n_{j,kl}^{{\rm sib}\pm \pm}
  -  \frac{1}{\epsilon_{\rm mix}} \sum_{j^{\prime} \neq j^{\prime \prime}}
\left[ \frac{\bar{N}^{\pm}}{n_{j^{\prime}}^{\pm}}
     + \frac{\bar{N}^{\pm}}{n_{j^{\prime \prime}}^{\pm}} - 1 \right]
n_{j^{\prime}k}^{\pm} n_{j^{\prime \prime}l}^{\pm}
\label{Eq39} \\
\Delta N^{\pm \mp}_{kl,\sigma-{\rm dyn}} & = &
\frac{1}{\epsilon} \sum_{j=1}^{\epsilon}
\frac{\bar{N}^{\pm}\bar{N}^{\mp}}{n_j^{\pm} n_j^{\mp}}
     n_{jk}^{\pm} n_{jl}^{\mp}
  -  \frac{1}{\epsilon_{\rm mix}} \sum_{j^{\prime} \neq j^{\prime \prime}}
\left[ \frac{\bar{N}^{\pm}}{n_{j^{\prime}}^{\pm}}
     + \frac{\bar{N}^{\mp}}{n_{j^{\prime \prime}}^{\mp}} - 1 \right]
n_{j^{\prime}k}^{\pm} n_{j^{\prime \prime}l}^{\mp}.
\label{Eq40}
\eea
\end{widetext}
Self-pair terms do not appear in $\sigma^2_{p_t,{\rm dynamical}}$. For like-sign sibling-pairs, events with $n_j^{\pm} = 1$, $n_{j,kl}^{{\rm sib}\pm \pm} = 0$ are skipped but those events are included in the mixed-event summation.

\subsection{Correlation derived from $F_{p_t}$}
\label{SecIIE}

Mean-$p_t$ fluctuation quantity $F_{p_t}$, developed by the PHENIX Collaboration~\cite{Fpt}, is based on a fluctuation width difference similar to $\Phi_{p_t}$. $F_{p_t}$ is defined by
\bea
F_{p_t} & = & \frac{\omega_{p_t,{\rm data}} - \omega_{p_t,{\rm mix}}}
                   {\omega_{p_t,{\rm mix}}}
\label{Eq41} \\
\omega_{p_t} & = & \left[ \overline{\langle p_t \rangle^2}
 - \overline{\langle p_t \rangle}^2 \right]^{1/2} / \overline{\langle p_t \rangle} ,
\label{Eq42}
\eea
where $\omega_{p_t}$ is calculated from the measured events (subscript ``data'') or from mixed events (subscript ``mix''). The latter are uncorrelated pseudo-events constructed by sampling from all particles in all events in the collection.

For mixed events Eq.~(\ref{Eq42}) becomes
\bea
\left( \overline{ \langle p_t \rangle}_{\rm mix} \omega_{p_t,{\rm mix}} \right)^2
 & = &
\overline{ \langle p_t \rangle^2_{\rm mix}} - 
\overline{ \langle p_t \rangle}^2_{\rm mix}
\nonumber \\
&   =  &
\overline{ \left( \langle p_t \rangle -  \overline{ \langle p_t \rangle}
           \right)^2_{\rm mix} }
\label{Eq43}
\eea
where
\bea
\overline{ \langle p_t \rangle}_{\rm mix} & = & 
\frac{1}{\epsilon^{\prime}} \sum_{j=1}^{\epsilon^{\prime}}
\frac{1}{n_j} \sum_{i=1}^{n_j} p_{t,ji}
   =
\sum_m \frac{\epsilon^{\prime}_m}{\epsilon^{\prime}}
\frac{1}{m \epsilon^{\prime}_m} 
\sum_{j=1}^{\epsilon^{\prime}_m}  \sum_{i=1}^m  p_{t,ji}
\nonumber \\
 & = & \sum_m \frac{\epsilon^{\prime}_m}{\epsilon^{\prime}} \hat{p}_{t,m}
= \hat{p}_t.
\label{Eq44}
\eea
In Eq.~(\ref{Eq44}) $\epsilon^{\prime}$ is the number of mixed events, $\epsilon^{\prime}_m$ is the number of mixed events having multiplicity $m$, $\hat{p}_{t,m}$ is the inclusive mean-$p_t$ for all mixed events with multiplicity $m$, and in the last step any systematic dependence on multiplicity of the inclusive mean-$p_t$ for the real events is suppressed because each mixed event is composed of a random sample of particles from all events in the collection. For real events, in which $\hat{p}_{t,m}$ may systematically vary with multiplicity, $\overline{ \langle p_t \rangle}_{\rm data} \neq \hat{p}_t$ in general. However, the ratio $\zeta = \hat{p}_t/\overline{ \langle p_t \rangle}_{\rm data}$ is expected to be approximately 1. Continuing from Eq.~(\ref{Eq43}) we obtain
\bea
\left( \hat{p}_t \omega_{p_t,{\rm mix}} \right)^2 
  &  = &
\frac{1}{\epsilon^{\prime}} \sum_{j=1}^{\epsilon^{\prime}}
\left( \langle p_t \rangle_j - \hat{p}_t \right)^2_{\rm mix}
\nonumber \\
 &  & \hspace{-0.8in} =  \sum_m \frac{\epsilon^{\prime}_m}{\epsilon^{\prime}}
\frac{1}{\epsilon^{\prime}_m}
\sum_{j=1}^{\epsilon^{\prime}_m}
\left( \langle p_t \rangle_{j,m} - \hat{p}_t \right)^2_{\rm mix}
\nonumber \\
 &  &  \hspace{-0.8in} = \sum_m \frac{\epsilon^{\prime}_m}{\epsilon^{\prime}}
\sigma^2_{\langle p_t \rangle :m,{\rm mix}}
  =
\sum_{m>0} \frac{\epsilon^{\prime}_m}{\epsilon^{\prime}}
\frac{\sigma_{\hat{p}_t}^2}{m} = \overline{m^{-1}} \sigma_{\hat{p}_t}^2
\label{Eq45}
\eea
where $\sigma^2_{\langle p_t \rangle :m,{\rm mix}}$ is the mean-$p_t$ variance for mixed events with multiplicity $m>0$ (the summation includes only those events with non-vanishing bin content), and in the last line the central limit theorem can be used because the mixed events are uncorrelated. Quantity $\overline{m^{-1}}$ is statistically biased and will be discussed below.

In order to access the correlation, the width difference form of $F_{p_t}$ must be transformed to a variance difference similar to what was done for $\Phi_{p_t}$ in Sec.~\ref{SecIIC}. This transformation can be accomplished by multiplying $F_{p_t}$ by $\bar{N}^2 \hat{p}_t^2 \omega_{p_t,{\rm mix}} (\omega_{p_t,{\rm data}} + \omega_{p_t,{\rm mix}})$. The result is defined with new symbol ${\cal F}_{p_t}$ where $\bar{N}^2$ is required in order that the resulting expression be proportional to pair number. 
In Ref.~\cite{Fpt} $F_{p_t}$ for RHIC collision data was found to be of order 2\% which implies that
\bea
\omega_{p_t,{\rm data}} & \approx & \omega_{p_t,{\rm mix}} =
\left[ \overline{ m^{-1} } \right] ^{1/2} \sigma_{\hat{p}_t} / \hat{p}_t
\label{Eq46}
\eea
using Eq.~(\ref{Eq45}). The above multiplicative factor is then approximately $2 \bar{N}^2 \overline{ m^{-1}} \sigma_{\hat{p}_t}^2$, which is a constant or scaling factor.

The result for ${\cal F}_{p_t}$ is given by
\bea
{\cal F}_{p_t} & = & \bar{N}^2 \hat{p}_t^2 \left( \omega^2_{p_t,{\rm data}}
- \omega^2_{p_t,{\rm mix}} \right)
\nonumber \\
 & = & \frac{\bar{N}^2 \hat{p}_t^2}{\overline{ \langle p_t \rangle}^2_{\rm data}}
\frac{1}{\epsilon} \sum_{j=1}^{\epsilon}
\left( \langle p_t \rangle_j - \overline{ \langle p_t \rangle} \right)^2_{\rm data}
- \bar{N}^2 \overline{ m^{-1}} \sigma_{\hat{p}_t}^2
\nonumber \\
 & = & \bar{N}^2 \zeta^2 \frac{1}{\epsilon} \sum_{j=1}^{\epsilon}
\left( \langle p_t \rangle_j - \hat{p}_t/\zeta \right)^2_{\rm data}
- \bar{N}^2 \overline{ m^{-1}} \sigma_{\hat{p}_t}^2 
\label{Eq47}
\eea
where the second quantity on the RHS was obtained from Eq.~(\ref{Eq45}).
If $\zeta \neq 1$, the statistical bias from the average factor $\overline{ m^{-1}}$ contributes to the final $\Delta N_{kl}$ correlation as an additive bias which may produce significant artifacts. It is expected that $\zeta \approx 1$ for applications in high energy heavy-ion collisions and therefore setting $\zeta = 1$ permits a formal, statistically unbiased correlation to be defined approximately corresponding to fluctuation quantity ${\cal F}_{p_t}$.

By evaluating Eq.~(\ref{Eq47}) for $(++)$ and $(--)$ charged-particle pairs and inserting the summations for $\langle p_t \rangle_j$, $\hat{p}_t$ and $\sigma_{\hat{p}_t}^2 $ as in the above derivations, the like-sign correlation quantity can be derived and is given by
\bea
\Delta N^{\pm \pm}_{kl,{\rm F}} & = & 
\frac{1}{\epsilon} \sum_{j=1}^{\epsilon} \frac{\bar{N}^{\pm^2}}
                     {n_j^{\pm^2}} n_{j,kl}^{{\rm sib} \pm \pm}
\nonumber \\
 & & \hspace{-0.8in} - \frac{1}{\epsilon_{\rm mix}} \sum_{j^{\prime} \neq j^{\prime \prime}}
\left( \frac{\bar{N}^{\pm}}{n^{\pm}_{j^{\prime}}}
  +     \frac{\bar{N}^{\pm}}{n^{\pm}_{j^{\prime \prime}}}
  - 1 - \overline{{m^{\pm}}^{-1}} \right)
n^{\pm}_{j^{\prime}k} n^{\pm}_{j^{\prime \prime}l}
\label{Eq48}
\eea
where self-pair terms are not included in the correlation. The statistical bias factor $ \overline{m^{\pm^{-1}}}$ in the mixed-event term is equal to $\bar{N}^{\pm^2} \overline{m^{\pm^{-1}}} \hat{\bar{n}}^{\pm}_k \hat{\bar{n}}^{\pm}_l$, if multiplicity dependence of the single-particle $p_t$ distribution shape is neglected. This bias term is cancelled by a similar bias term in the sibling-pair sum given by
\bea
\bar{N}^{\pm^2} \overline{ \left( \frac{m^{\pm} (m^{\pm} - 1)}{{m^{\pm}}^2} \right)  }
\hat{\bar{n}}^{{\rm sib}\pm \pm}_{kl}
 & = &
\bar{N}^{\pm^2} \left( 1 - \overline{m^{\pm^{-1}}} \right)
\hat{\bar{n}}^{{\rm sib}\pm \pm}_{kl}
\nonumber \\
 &  &  \hspace{-0.5in} \approx  
\bar{N}^{\pm^2} \left( 1 - \overline{m^{\pm^{-1}}} \right)
\hat{\bar{n}}^{\pm}_k \hat{\bar{n}}^{\pm}_l
\label{Eq49}
\eea
where the bias contribution comes from the second term on the RHS. Multiplicity dependence in the shape of the two-particle distribution is also neglected in Eq.~(\ref{Eq49}). The last line in Eq.~(\ref{Eq49}) represents the limit of zero correlations. For realistic applications with non-vanishing correlations this statistical bias contributes to $\Delta N^{\pm \pm}_{kl,{\rm F}}$. In Sec.~\ref{SecV} the possible significance of this bias will be studied using simulations.

For unlike-sign pairs quantity ${\cal F}_{p_t}$, with $\zeta = 1$, becomes
\bea
{\cal F}_{p_t}^{\rm US} & = & \frac{\bar{N}^+ \bar{N}^-}{\epsilon}
\sum_{j=1}^{\epsilon}
\left( \langle p_t^+ \rangle_j - \hat{p}_t^+ \right)
\left( \langle p_t^- \rangle_j - \hat{p}_t^- \right).
\label{Eq50}
\eea
The resulting unlike-sign correlation is given by
\bea
\Delta N^{\pm \mp}_{kl,{\rm F}} & = &
\frac{1}{\epsilon} \sum_{j=1}^{\epsilon}
\frac{\bar{N}^+ \bar{N}^-}{n_j^+ n_j^-} 
n_{j,kl}^{{\rm sib} \pm \mp}
\nonumber \\
  & & \hspace{-0.4in} - \frac{1}{\epsilon_{\rm mix}} \sum_{j^{\prime} \neq j^{\prime \prime}}
\left( \frac{\bar{N}^{\pm}}{n^{\pm}_{j^{\prime}}}
  +     \frac{\bar{N}^{\mp}}{n^{\mp}_{j^{\prime \prime}}} -1 \right)
n^{\pm}_{j^{\prime}k} n^{\mp}_{j^{\prime \prime}l}
\label{Eq51}
\eea
which is statistically unbiased.

\subsection{Correlations derived from $\Delta [P_T,N]$ and $\Sigma [P_T,N]$}
\label{SecIIF}

The NA49 Collaboration recently published transverse momentum and multiplicity fluctuation measures $\Delta [P_T,N]$ and $\Sigma [P_T,N]$~\cite{NA49}, defined in the present notation by
\bea
\Delta [P_T,N] & = & \frac{1}{\bar{N} \omega(p_t)} 
\left\{ \bar{N} \omega[P_T] - \overline{P}_T \omega[N] \right\},
\label{Eq52} \\
\Sigma [P_T,N] & = & \frac{1}{\bar{N} \omega(p_t)}
\left\{ \bar{N} \omega[P_T] + \overline{P}_T \omega[N] \right. 
\nonumber \\
 &  - & \left. 2\left[ \overline{P_T N} - \overline{P}_T \bar{N} \right] \right\}.
\label{Eq53}
\eea
In these equations $P_{T,j} = \sum_{i=1}^{n_j} p_{t,ji}$ is the event-wise sum of $p_t$ magnitude over all particles in the $j^{\rm th}$ event. The other symbols are defined as follows:
\bea
\overline{P}_T & = & \frac{1}{\epsilon} \sum_{j=1}^{\epsilon} P_{T,j},
\label{Eq54} \\
\overline{P_T N} & = & \frac{1}{\epsilon} \sum_{j=1}^{\epsilon} n_j P_{T,j},
\label{Eq55} \\
 \omega[P_T] & = & \left( \overline{P_T^2} - \overline{P}_T^2 \right) /\overline{P}_T,
\label{Eq56} \\
\omega[N] & = & \left( \overline{N^2} - \bar{N}^2 \right) / \bar{N},
\label{Eq57} \\
\omega(p_t) & = & \sigma^2_{\hat{p}_t} / \hat{p}_t .
\label{Eq58}
\eea
Note the different meaning of symbol $\omega$ in the above equations from Ref.~\cite{NA49}, which is proportional to a fluctuation variance, compared to the definition in the previous subsection where that $\omega$ was proportional to a fluctuation width.

Equation~(\ref{Eq52}) can be simplified by multiplying both numerator and denominator of the RHS by $\bar{N}\overline{P}_T$. The result is
\bea
\overline{P}_T \left( \sigma^2_{\hat{p}_t} / \hat{p}_t \right) \Delta [P_T,N]
 & = &
\left[ \bar{N}^2 \overline{P_T^2} - \overline{N^2} \, \overline{P}_T^2
\right] / \bar{N}^2 .
\nonumber \\
\label{Eq59}
\eea
The correlated particle-pair difference is derived by inserting summations in the numerator of the RHS of this equation, omitting a self-pair term, and binning on 2D transverse momentum. The result is given by
\bea
\Delta N_{kl,\Delta} & = & \frac{1}{\epsilon} \sum_{j=1}^{\epsilon}
n^{\rm sib}_{j,kl}
- \frac{\overline{N^2}}{\bar{N}^2} \frac{1}{\epsilon_{\rm mix}}
\sum_{j \neq j^{\prime}} n_{jk} n_{j^{\prime}l} .
\label{Eq60}
\eea
The multiplicity bin-width dependence of $\Delta N_{kl,\Delta}$ can be estimated using the same steps as in Eq.~(\ref{Eq7}) and neglecting possible systematic variations in the shapes of the single- and two-particle distributions on $p_t$ with event multiplicity. The result is
\bea
\Delta N_{kl,\Delta} & \approx &
\left[ \bar{N} (\bar{N} - 1) + \sigma^2_N \right] \hat{\bar{n}}^{\rm sib}_{kl}
- \left( \bar{N}^2 + \sigma^2_N \right) \hat{\bar{n}}_k \hat{\bar{n}}_l
\nonumber \\
 & = & \left( \bar{N}^2 + \sigma^2_N \right) \left( \hat{\bar{n}}^{\rm sib}_{kl} - \hat{\bar{n}}_k \hat{\bar{n}}_l \right) -  \bar{N} \hat{\bar{n}}^{\rm sib}_{kl} .
\label{Eq61}
\eea
The last term on the RHS of Eq.~(\ref{Eq61}) represents an additive bias, i.e. $\Delta N_{kl,\Delta} \neq 0$ in the no-correlation limit, $\hat{\bar{n}}^{\rm sib}_{kl} = \hat{\bar{n}}_k \hat{\bar{n}}_l$.

The correlated particle-pair difference for $\Sigma [P_T,N]$ can be derived by following the same steps as above, assuming large event-number $\epsilon >\!> 1$. The result is given by
\bea
\Delta N_{kl,\Sigma} & = & \frac{1}{\epsilon} \sum_{j=1}^{\epsilon}
n^{\rm sib}_{j,kl}
 -  \frac{1}{\epsilon_{\rm mix}} \sum_{j \neq j^{\prime}}
\left( \frac{2n_{j^{\prime}}}{\bar{N}} - \frac{\overline{N^2}}{\bar{N}^2} \right) n_{jk} n_{j^{\prime}l}
\nonumber \\
\label{Eq62} \\
 & \approx & \left[ \bar{N}(\bar{N}-1)+\sigma^2_N \right] \hat{\bar{n}}^{\rm sib}_{kl}
-2\left( \bar{N}^2 + \sigma^2_N \right)\hat{\bar{n}}_k \hat{\bar{n}}_l
\nonumber \\
 & + &  \left( \bar{N}^2 + \sigma^2_N \right)\hat{\bar{n}}_k \hat{\bar{n}}_l
\nonumber \\
 & = & \left( \bar{N}^2 + \sigma^2_N \right) \left( \hat{\bar{n}}^{\rm sib}_{kl} - \hat{\bar{n}}_k \hat{\bar{n}}_l \right) -  \bar{N} \hat{\bar{n}}^{\rm sib}_{kl} 
\label{Eq63}
\eea
where the second and third equations apply if the shapes of the single- and two-particle $p_t$-distributions do not vary with event multiplicity. In this limit $\Delta N_{kl,\Delta}$ and $\Delta N_{kl,\Sigma}$ are equal and both are additively biased. The bias can be minimized by selecting multiplicity ranges where $\sigma^2_N \ll \bar{N}^2$ and multiplying the mixed-event pair term with factor $(\bar{N}-1)/\bar{N}$, or by selecting multiplicity ranges where $\sigma^2_N \approx \bar{N}$ (Poisson limit) and $\bar{N} \gg \sigma_N$.

\subsection{Multiplicity dependent non-statistical fluctuations}
\label{SecIIFb}

In the preceding subsections two-particle correlations on $(p_{t1},p_{t2})$ were derived from different non-statistical mean-$p_t$ fluctuation quantities. For example, $\Delta \sigma^2_{p_t:m}$ in Eq.~(\ref{Eq15}) is expressed as an average over all events in the centrality bin, given by
\bea
\Delta \sigma^2_{p_t:m} & = & \overline{m \sigma^2_{\langle p_t \rangle} - \sigma^2_{\hat{p}_t}}
\label{Eq63a}
\eea
where $m$ is the $(\eta,\phi)$ bin-wise multiplicity and $\sigma^2_{\langle p_t \rangle}$ is the variance of fluctuating mean-$p_t$ for the events containing $m$ particles. Similar expressions for the non-charge-identified forms for $\Phi_{p_t}^{(0)}$, $\sigma^2_{p_t,{\rm dynamical}}$ and ${\cal F}_{p_t}$ are given by
\bea
2 \sigma_{\hat{p}_t} \Phi_{p_t}^{(0)} & = & \overline{m ( m \sigma^2_{\langle p_t \rangle} - \sigma^2_{\hat{p}_t})}/\bar{N}
\label{Eq63b} \\
(\bar{N} - 1) \sigma^2_{p_t,{\rm dynamical}} & = & (\bar{N} - 1)
\overline{\left( \frac{m \sigma^2_{\langle p_t \rangle} - \sigma^2_{\hat{p}_t}}{m-1}\right)}
\label{Eq63c} \\
{\cal F}_{p_t}/\bar{N} & \approx & \bar{N} \overline{( m \sigma^2_{\langle p_t \rangle} - \sigma^2_{\hat{p}_t})/m}
\label{Eq63d}
\eea
using Eqs.~(\ref{Eq32}), (\ref{Eq37}) and (\ref{Eq47}) and rearranging the event summation as in Eq.~(\ref{Eq15}). In the last equation $\zeta \approx 1$ was assumed for quantity ${\cal F}_{p_t}$.

Non-statistical fluctuations, or two-particle correlations, cause $( m \sigma^2_{\langle p_t \rangle} - \sigma^2_{\hat{p}_t}) \neq 0$ and to depend on $(\eta,\phi)$ bin-wise multiplicity. Defining
\bea
f(m) & \equiv & m \sigma^2_{\langle p_t \rangle} - \sigma^2_{\hat{p}_t}
\label{Eq63e}
\eea
each quantity can be written as a multiplicity weighted average of $f(m)$, given by
\bea
\Delta \sigma^2_{p_t:m} & = & \overline{f(m)}
\label{Eq63f} \\
2 \sigma_{\hat{p}_t} \Phi_{p_t}^{(0)} & = & \overline{mf(m)}/\bar{N}
\label{Eq63g} \\
(\bar{N} - 1) \sigma^2_{p_t,{\rm dynamical}} & = & (\bar{N} - 1)
\overline{ f(m)/(m-1)}
\label{Eq63h} \\
{\cal F}_{p_t}/\bar{N} & \approx & \bar{N} \overline{f(m)/m}
\label{Eq63i}
\eea
where the explicit dependence of $f(m)$ on multiplicity is determined by the dynamical processes which produce non-statistical fluctuations. For example, event-wise fluctuations in global temperature involve all particles such that the number of correlated particle-pairs, $\Delta{N}_{kl}$, is proportional to $m^2$ and $\Delta \sigma^2_{p_t:m}$, being proportional to number of correlated pairs per final-state particle [see Eq.~(\ref{Eq19})], is proportional to $m$. For this example $f(m) \propto m$ and the averages over finite width multiplicity bins in above Eqs.~(\ref{Eq63f}) - (\ref{Eq63i}) result in a bias (results depend on bin width) for $\Phi_{p_t}^{(0)}$ but not for the other three quantities. The bias in $\Phi_{p_t}^{(0)}$ occurs because $\overline{mf(m)} = \overline{m^2} = \bar{N}^2 + \sigma^2_N$. If the number of correlated particle pairs is proportional to $m$, then $f(m)$ is constant and quantities $\sigma^2_{p_t,{\rm dynamical}}$ and ${\cal F}_{p_t}$ are biased while $\Delta \sigma^2_{p_t:m}$ and $\Phi_{p_t}^{(0)}$ are not.

If an analysis of data were focused on a specific dynamical process which was known to produce a certain $f(m)$, then the set of possible correlation quantities could be ranked with respect to optimal suppression of the above bias effect. For practical analysis of data from relativistic heavy-ion collisions multiple dynamical processes contribute to the non-statistical fluctuations and those processes are expected to follow different functions of multiplicity, for example the number of nucleon participants or the number of binary nucleon-nucleon collisions~\cite{axialCI}. Dynamical processes also depend on the charge combination of particle-pairs (e.g. in hadronization) and the location in $(p_{t1},p_{t2})$ space (e.g. for soft versus semi-hard processes). Given this complexity it is preferable to evaluate the bias caused by multiplicity dependent non-statistical fluctuations, or correlations, by using realistic estimates of those correlations on $(p_{t1},p_{t2})$. This is the approach followed here and discussed in detail in Secs.~\ref{SecIII} and \ref{SecIV}.

\subsection{Normalized covariance}
\label{SecIIG}

In the preceding subsections the bin-wise number of correlated particle pairs $\Delta N_{kl}$ was calculated. In terms of particle densities $\Delta N_{kl}$ is proportional to $C(p_{t1},p_{t2})$ in Eq.~(\ref{Eq3}) which can be expressed as
\bea
\hat\rho (p_{t1},p_{t2}) & = & \hat\rho(p_{t1}) \hat\rho(p_{t2}) +
C(p_{t1},p_{t2})
\nonumber \\
 &  & \hspace{-0.6in} =  \hat\rho(p_{t1}) \hat\rho(p_{t2}) r(p_{t1},p_{t2})
\nonumber \\
 &  & \hspace{-0.6in} = \hat\rho(p_{t1}) \hat\rho(p_{t2}) \left\{ 1 +\left[ r(p_{t1},p_{t2}) - 1 \right] \right\},
\label{Eq64}
\eea
where the correlated pair density is given by
\bea
C(p_{t1},p_{t2}) & = & \hat\rho(p_{t1}) \hat\rho(p_{t2})\left[ r(p_{t1},p_{t2}) - 1 \right].
\nonumber \\
\label{Eq65}
\eea
Quantities $C$ and $\Delta N_{kl}$ therefore include a trivial dependence on multiplicity squared which is easily removed by dividing by $\hat\rho(p_{t1}) \hat\rho(p_{t2})$. Furthermore, tracking inefficiency and detector acceptance effects cancel in this ratio if the product $\hat\rho(p_{t1}) \hat\rho(p_{t2})$ is calculated using the same data which were used for $C(p_{t1},p_{t2})$.

In heavy-ion collisions, quantum interference between identical particles in the final-state produces correlations which scale with the number of identical-particle pairs~\cite{HBT}. The per-pair ratio $C(p_{t1},p_{t2})/\hat\rho(p_{t1}) \hat\rho(p_{t2})$ is approximately constant with increasing centrality. All other processes which are expected to produce correlations (see Sec.~\ref{SecI}) scale with either the number of participating nucleons, the number of binary nucleon + nucleon collisions, or the number of final-state particles. A per final-state particle ratio~\cite{Trainormeanpt} is therefore more appropriate for studying the centrality dependence of most correlation structures other than that caused by final-state quantum interference.

In the statistics literature Pearson's normalized covariance~\cite{Pearson}, given by
\bea
\frac{\overline{(n_k - \bar{n}_k)(n_l - \bar{n}_l)}}
     {\sqrt{\sigma^2_{n_k} \sigma^2_{n_l}}}
 & = &
\frac{\overline{n_k n_l} - \bar{n}_k \bar{n}_l}{\sigma_{n_k} \sigma_{n_l}}
\nonumber \\
 & \approx &  \frac{\overline{n_k n_l} - \bar{n}_k \bar{n}_l}{\sqrt{\bar{n}_k \bar{n}_l}},
\label{Eq66}
\eea
provides the necessary, per final-state particle correlation measure using the geometric-mean particle number in the denominator, where $n_k$ and $n_l$ are the event-wise number of particles in bins $k$ and $l$. In Eq.~(\ref{Eq66}) over-lines indicate event averages within the event collection, and $\sigma_{n_k}^2$ is the variance of the event-wise, multiplicity frequency distribution in bin $k$. In the last step in Eq.~(\ref{Eq66}) the bin-wise multiplicity frequency distributions are assumed to be accurately represented by Poisson distributions with means $\bar{n}_k$ and $\bar{n}_l$. In order to cancel efficiency and acceptance effects, applications of Eq.~(\ref{Eq66})~\cite{axialCI} take the form
\bea
\frac{\overline{n_k n_l} - \bar{n}_k \bar{n}_l}{\sqrt{\bar{n}_k \bar{n}_l}}
 & = & \sqrt{\bar{n}_k \bar{n}_l} \left[ \frac{\overline{n_k n_l} - \bar{n}_k \bar{n}_l}{\bar{n}_k \bar{n}_l} \right]
\label{Eq67}
\eea
where the ratio in square brackets is obtained from the data and the leading square-root, or prefactor (see Sec.~\ref{SecIIA}), is calculated from a model representation of the efficiency corrected product of single-particle yields. For transverse momentum correlations the prefactor $\sqrt{\rho_{\rm ref}}$ is given by
\bea
\sqrt{\rho_{\rm ref}(p_{t1},p_{t2})} & = & \left[ \frac{d^2N_{\rm ch}}{dp_{t1} d\eta_1}
                 \frac{d^2N_{\rm ch}}{dp_{t2} d\eta_2} \right] ^{1/2}
\label{Eq68}
\eea
where $N_{\rm ch}$ includes all charged particles within the $p_t$, $\Delta\eta$ and $\Delta\phi$ acceptance.

Parton fragmentation into jets is of considerable interest in analysis of heavy-ion collision data. It has been shown that jet fragment distributions scale with transverse  rapidity $y_t$~\cite{Tomjetfrag}, defined by
\bea
y_t & = & \ln [(p_t + m_t)/m_0],
\label{Eq69}
\eea
where $m_t = \sqrt{p_t^2 + m_0^2}$, and arbitrary mass $m_0$, which regulates the singularity at $p_t = 0$, is assumed equal to the pion mass when non-identified particles are used in the analysis, and equals the true particle-mass when the particle species is identified. The single-particle distribution on $y_t$ is given by
\bea
\frac{d^2N_{\rm ch}}{dy_t d\eta} & = & \frac{dp_t}{dy_t}
\frac{d^2N_{\rm ch}}{dp_t d\eta} \approx m_t \frac{d^2N_{\rm ch}}{dp_t d\eta}
\label{Eq70}
\eea
where $p_t = m_0\sinh (y_t)$ and in the last step $dp_t/dy_t = m_t$ at mid-rapidity ($\eta = 0$) is assumed. In the present application 2D transverse rapidity correlations will be calculated where the final quantity, $\Delta\rho/\sqrt{\rho_{\rm ref}}$, is given by
\bea
\frac{\Delta\rho}{\sqrt{\rho_{\rm ref}}} (y_{t,k},y_{t,l})
 & = &
\left[ \frac{d^2N_{\rm ch}}{dy_{t,k} d\eta}
                 \frac{d^2N_{\rm ch}}{dy_{t,l} d\eta} \right] ^{1/2}
\frac{\Delta N_{kl}}{N^{\rm mix}_{kl}}.
\label{Eq71}
\eea
Pair difference $\Delta N_{kl}$ for like-sign and unlike-sign particle pairs for all the various methods derived here are given in the preceding subsections. The mixed-event particle pair averages $N^{\rm mix}_{kl}$ are given by the second factor on the right-hand sides of Eqs.~(\ref{Eq21}), (\ref{Eq23}), (\ref{Eq26}), (\ref{Eq34}), (\ref{Eq35}), (\ref{Eq39}), (\ref{Eq40}), (\ref{Eq48}) and (\ref{Eq51}) for each mean-$p_t$ fluctuation quantity considered here. Sibling-pair averages $N^{\rm sib}_{kl}$ are given by the first factors on the right-hand sides in the preceding list of equations.

The prefactor in Eqs.~(\ref{Eq68}) and (\ref{Eq71}) applies when all charged-particle pairs in the acceptance are used in the correlations. The number of like-sign or unlike-sign pairs is one-half of the total, assuming the numbers of positive and negative charged-particles are equal, which is approximately true for relativistic heavy-ion collisions~\cite{STARspectra,PHENIXspectra,ALICEspectra}. The appropriate prefactor for these cases is $\sqrt{\rho_{\rm ref}/2}$.

In Sec.~\ref{SecIV} Monte Carlo simulations are described for  each mean-$p_t$ fluctuation quantity in which LS, US, charge-independent (CI) and charge-dependent (CD) correlations, plus alternate CI and CD forms are included. For each correlation form, charged-particle pair combinations $(a,b) = (++)$, $(--)$, $(+-)$ and $(-+)$ are calculated and combined to give the LS, US, CI and CD combinations.  Those combinations and the corresponding prefactors are listed in the following equations:
\bea
\frac{\Delta\rho}{\sqrt{\rho_{\rm ref}}} ({\rm LS})   & = &
\frac{\sqrt{\rho_{\rm ref}}}{2\sqrt{2}} 
\sum_{ab=++,--}  \left[ \frac{\Delta N_{kl}}{N^{\rm mix}_{kl}} \right]_{ab}
\label{Eq72} \\
\frac{\Delta\rho}{\sqrt{\rho_{\rm ref}}} ({\rm US})   & = &
\frac{\sqrt{\rho_{\rm ref}}}{2\sqrt{2}}
\sum_{ab=+-,-+}  \left[ \frac{\Delta N_{kl}}{N^{\rm mix}_{kl}} \right]_{ab}
\label{Eq73} \\
\frac{\Delta\rho}{\sqrt{\rho_{\rm ref}}} ({\rm CI})   & = &
\frac{\sqrt{\rho_{\rm ref}}}{4}
\sum_{a,b=\pm,\pm}  \left[ \frac{\Delta N_{kl}}{N^{\rm mix}_{kl}} \right]_{ab}
\label{Eq74} \\
\frac{\Delta\rho}{\sqrt{\rho_{\rm ref}}} ({\rm CI,alt})   & = &
\sqrt{\rho_{\rm ref}}
\frac{\sum_{a,b=\pm,\pm} \Delta N^{ab}_{kl}}
     {\sum_{a,b=\pm,\pm} N^{{\rm mix},ab}_{kl}}
\label{Eq75} \\
\frac{\Delta\rho}{\sqrt{\rho_{\rm ref}}} ({\rm CD})   & = &
\nonumber \\
 & & \hspace{-0.9in} \sqrt{\rho_{\rm ref}}
\frac{  \left( N^{\rm sib,++}_{kl} + N^{\rm sib,--}_{kl} \right)
      - \left( N^{\rm sib,+-}_{kl} + N^{\rm sib,-+}_{kl} \right) }
     {\sum_{a,b=\pm,\pm} N^{{\rm mix},ab}_{kl}}
\nonumber \\
\label{Eq76} \\
\frac{\Delta\rho}{\sqrt{\rho_{\rm ref}}} ({\rm CD,alt})   & = &
\nonumber \\
 & & \hspace{-0.9in}
\sqrt{\rho_{\rm ref}}
\frac{  \left( \Delta N_{kl}^{++} + \Delta N_{kl}^{--} \right)
      - \left( \Delta N_{kl}^{+-} + \Delta N_{kl}^{-+} \right) }
     {\sum_{a,b=\pm,\pm} N^{{\rm mix},ab}_{kl}}
\nonumber \\
\label{Eq77}
\eea
where summation indices $a,b$ denote charged-particle pair combinations.

\section{Systematic bias}
\label{SecIII}

In addition to the pair-counting statistical bias caused by finite centrality bin widths, systematic variations in the shapes of the single-particle distribution and the true correlation itself within the range of the centrality bin lead to bias in the correlation quantities. Systematic bias due to shape variation in the single-particle distribution occurs because: (1) mixed-event particle pairs are selected from arbitrary pairs of events within the centrality bin where each mixed-event particle pair may sample two different parent distributions;~\footnote{A parent distribution is the infinite statistics limit of a measured distribution which in this analysis is approximated by a mathematical function. Finite statistics random samples of the parent distribution produce fluctuating event-wise distributions.} (2) sibling pairs from a single-event sample the same parent distribution, while sibling pairs for other events may sample different parent distributions; (3) ``cross-terms'' in the mixed-event pairs, where different parent distributions are sampled, have no corresponding terms in the sibling pairs, giving rise to non-vanishing contributions in the absence of true correlations. 

Systematic bias due to multiplicity dependent shape variations in the true correlation is a matter of definition. Here, the goal is to measure the true correlation at a fixed multiplicity or centrality. The amount by which the measured correlation, when averaged over the finite centrality bin-width, differs from the true correlation at the mid-point of the bin is considered a bias. On the other hand, if the goal is to measure the average correlation within the finite centrality bin then the bias, if any, will depend on the averaging method. The effects of these sources of systematic bias will be discussed in Sec.~\ref{SecIV} in terms of Monte Carlo simulations.

In this section we illustrate the origin of systematic bias by calculating systematic contributions to the $\Delta\sigma^2_{p_t:m}$ based correlation to leading order, where for simplicity, charge identification is ignored. The purpose of the analytical treatment in this section is to provide an intuitive understanding of the above two sources of systematic bias.
From Eq.~(\ref{Eq21}) the sibling-pair number can be written as
\bea
N^{\rm sib}_{kl,\Delta\sigma^2} & = &
\frac{1}{\epsilon} \sum_{j=1}^{\epsilon}
\frac{\bar{N}}{n_j}
n^{\rm sib}_{j,kl}
=
\sum_m \frac{\epsilon_m}{\epsilon} \frac{\bar{N}}{m} \frac{1}{\epsilon_m}
\sum_{j=1}^{\epsilon_m}
n^{\rm sib}_{jm,kl}
\nonumber \\
 & = & \sum_m \frac{\epsilon_m}{\epsilon} \frac{\bar{N}}{m}
\bar{n}^{\rm sib}_{m,kl}
=
 \sum_m \frac{\epsilon_m}{\epsilon} \frac{\bar{N}}{m} m(m-1)
\hat{\bar{n}}^{\rm sib}_{m,kl}
\nonumber \\
\label{Eq78}
\eea
where, as in Eq.~(\ref{Eq7}), $m$ is a multiplicity within the finite bin, $\bar{N}$ is the mean multiplicity, $\epsilon_m$ is the number of events having multiplicity $m$, $\bar{n}^{\rm sib}_{m,kl}$ is the average sibling-pair histogram for all $\epsilon_m$ events, and $\hat{\bar{n}}^{\rm sib}_{m,kl}$ is the unit-normalized distribution on $y_t$ bins $(k,l)$ where $\sum_{k,l} \hat{\bar{n}}^{\rm sib}_{m,kl} = 1$.
The sibling-pair distribution consists of an uncorrelated (factorized) part plus a non-factorized correlation, which is written for the unit-normalized distributions as
\bea
\hat{\bar{n}}^{\rm sib}_{m,kl} & = & \hat{\bar{n}}_{mk} \hat{\bar{n}}_{ml}
+ C_{m,kl}
\label{Eq79}
\eea
where $\hat{\bar{n}}_{mk}$ is the average, unit-normalized single-particle distribution on $y_t$ bin $k$ for events having multiplicity $m$, where $\sum_k \hat{\bar{n}}_{mk} = 1$. True correlation quantity $C_{m,kl}$ is defined such that $\sum_{kl}C_{m,kl} = 0$.

We consider the possibility that the shapes of both the single-particle distribution and the true correlation vary with multiplicity $m$, and therefore express these quantities as
\bea
\hat{\bar{n}}_{mk} & = & \hat{\bar{n}}_{k}^{(0)} + \delta \hat{\bar{n}}_{mk},
\label{Eq80} \\
C_{m,kl} & = & C_{kl}^{(0)} + \delta C_{m,kl}
\label{Eq81}
\eea
where $\hat{\bar{n}}_{k}^{(0)}$ is the single-particle distribution at the mid-point of the bin and $C_{kl}^{(0)}$ is the true correlation at a fixed multiplicity which is the primary object to be measured. Substituting Eqs.~(\ref{Eq79}) - (\ref{Eq81}) into Eq.~(\ref{Eq78}) yields
\begin{widetext}
\bea
N^{\rm sib}_{kl,\Delta\sigma^2} & = & \bar{N}(\bar{N}-1)
\left[ \hat{\bar{n}}_{k}^{(0)} \hat{\bar{n}}_{l}^{(0)} + C_{kl}^{(0)} \right]
  +   
\bar{N} \sum_m \frac{\epsilon_m}{\epsilon} (m-1) \left[
\hat{\bar{n}}_{k}^{(0)} \delta \hat{\bar{n}}_{ml} +
\hat{\bar{n}}_{l}^{(0)} \delta \hat{\bar{n}}_{mk}
  +  \delta \hat{\bar{n}}_{mk} \delta \hat{\bar{n}}_{ml} +
\delta C_{m,kl} \right].
\label{Eq82}
\eea
Similarly the mixed-event pair distribution from Eq.~(\ref{Eq21}) is expressed as
\bea
N^{\rm mix}_{kl,\Delta\sigma^2} & = & \bar{N}(\bar{N}-1)
 \hat{\bar{n}}_{k}^{(0)} \hat{\bar{n}}_{l}^{(0)}
  +  (\bar{N}-1)  \sum_m \frac{\epsilon_m}{\epsilon} m
\left( \hat{\bar{n}}_{k}^{(0)} \delta \hat{\bar{n}}_{ml}
     + \hat{\bar{n}}_{l}^{(0)} \delta \hat{\bar{n}}_{mk} \right)
\nonumber \\
 & +  &   \frac{\bar{N}-1}{\bar{N}}
\left(  \sum_m \frac{\epsilon_m}{\epsilon} m \delta \hat{\bar{n}}_{mk} \right)
\left(  \sum_{m^{\prime}} \frac{\epsilon_{m^{\prime}}}{\epsilon} {m^{\prime}} \delta \hat{\bar{n}}_{{m^{\prime}}l} \right).
\label{Eq83}
\eea
The correlated pair-difference is given by
\bea
\Delta N_{kl,\Delta\sigma^2} & = & \bar{N}(\bar{N}-1) C_{kl}^{(0)}
+ \sum_m \frac{\epsilon_m}{\epsilon} (m-\bar{N})
\left( \hat{\bar{n}}_{k}^{(0)} \delta \hat{\bar{n}}_{ml} +
       \hat{\bar{n}}_{l}^{(0)} \delta \hat{\bar{n}}_{mk} \right)
\nonumber \\
 & + & \sum_m \frac{\epsilon_m}{\epsilon} \bar{N} (m-1)
\left( \delta \hat{\bar{n}}_{mk} \delta \hat{\bar{n}}_{ml} + \delta C_{m,kl} \right)
-  \frac{\bar{N}-1}{\bar{N}}
\left(  \sum_m \frac{\epsilon_m}{\epsilon} m \delta \hat{\bar{n}}_{mk} \right)
\left(  \sum_{m^{\prime}} \frac{\epsilon_{m^{\prime}}}{\epsilon} {m^{\prime}} \delta \hat{\bar{n}}_{{m^{\prime}}l} \right).
\label{Eq84}
\eea
\end{widetext}

Leading-order estimates of the bias contributions in Eq.~(\ref{Eq84}) are calculated as follows. We define $\delta_m \equiv m - \bar{N}$ as in Eq.~(\ref{Eq7}) and expand the change in shape of the single-particle distribution as
\bea
 \delta \hat{\bar{n}}_{mk} & \approx & \frac{\partial \hat{\bar{n}}_{mk}}
{\partial m} \mid _{m=\bar{N}} \delta_m + \cdots
\nonumber \\
 & \equiv & f_k \delta_m + \cdots ,
\label{Eq85}
\eea
and the change in shape of the true correlations as
\bea
\delta C_{m,kl} & = & \frac{\partial C_{m,kl}}
{\partial m} \mid _{m=\bar{N}} \delta_m + \cdots
\nonumber \\
 & \equiv & g_{kl} \delta_m + \cdots ~.
\label{Eq86}
\eea
The first-order expansion for the correlated pair-difference is derived by substituting the leading terms in Eqs.~(\ref{Eq85}) and (\ref{Eq86}) into Eq.~(\ref{Eq84}), replacing $m$ with $(\bar{N} + \delta_m)$, and retaining terms only through order $(\delta_m)^2$. The final result is given by
\bea
\Delta N_{kl,\Delta\sigma^2} & \approx & \bar{N}(\bar{N}-1) C_{kl}^{(0)}
+ \left( f_l  \hat{\bar{n}}_{k}^{(0)} + f_k \hat{\bar{n}}_{l}^{(0)} \right)
\sigma^2_N
\nonumber \\
 &  & \hspace{-0.5in} + \bar{N}(\bar{N}-1) f_k f_l \sigma^2_N
\left( 1 - \sigma^2_N / \bar{N}^2 \right) + \bar{N} g_{kl} \sigma^2_N
\label{Eq87}
\eea
where $\sigma^2_N = \sum_m (\epsilon_m/\epsilon) (\delta_m)^2$ is the variance of the finite width multiplicity bin. In the limit of zero multiplicity bin width quantity $\Delta N_{kl,\Delta\sigma^2}$ is proportional to the true correlation at a specific multiplicity. However, in general, Eq.~(\ref{Eq87}) shows that linear variations alone in the single-particle distribution and/or correlation shapes are sufficient to produce systematic bias. This bias occurs for the LS and US forms for all correlation quantities in Sec.~\ref{SecII}. In the next section Monte Carlo simulations are used to study a variety of realistic systematic variations in the shapes of the input distributions. These variations are general and are not limited to the linear terms assumed in this section.

\section{Monte Carlo simulations}
\label{SecIV}

Simulations were done for the correlation quantities derived in Sec.~\ref{SecII} using realistic multiplicity frequency distributions, centrality-bin widths, charged-particle $p_t$ spectra, and transverse momentum correlations in order to estimate realistic levels of statistical and systematic biases. The simulated collision system is minimum-bias (unrestricted, random nucleus + nucleus impact parameters) Au + Au collisions at $\sqrt{s_{NN}}$ = 200~GeV per colliding nucleon + nucleon pair. Collisions were selected for peripheral, mid-central and more-central nuclear overlap corresponding to total reaction cross section ranges~\cite{axialCI} 84-93\%, 55-64\% and 9-18\%, respectively, where, for example, the 5\% most-central (head-on) collisions account for the 0-5\% range of the multiplicity frequency distribution. Charged-particle production for $p_t > 0.15$~GeV/$c$, $|\eta| \leq 1$ and full $2\pi$ range in azimuth was assumed corresponding to the acceptance of the STAR TPC~\cite{STARTPC}. Monte Carlo Glauber model~\cite{MCG} estimates of the number of nucleon participants ($N_{\rm part}$) were taken from Ref.~\cite{axialCI}.

The minimum-bias multiplicity frequency distribution for relativistic heavy-ion collisions is accurately approximated by the power-law function~\cite{MCG}
\bea
\frac{dN_{\rm events}}{dN_{\rm ch}} & \propto & N_{\rm ch}^{-3/4}
\label{Eq88}
\eea
except near the lower and upper multiplicity end-points where the measured distribution falls off rapidly. For the selected centralities the multiplicity ranges are [2, 14], [66, 117] and [644, 910] with mean multiplicities $\bar{N}_{\rm ch}$ = 6.6, 89.7 and 771, respectively~\cite{MikeThesis}. Measurements~\cite{Prabhat} of the frequency distribution on charge difference $n_{\Delta} \equiv n^+ - n^-$ for positive $(n^+)$ and negative $(n^-)$ charged-particle multiplicities ($N_{\rm ch} = n^+ + n^-$) for all centralities indicate an approximate Gaussian dependence, $\exp[-(n_{\Delta} - \bar{n}_{\Delta})^2/2\sigma^2_{n_{\Delta}}]$, where the mean $(\bar{n}_{\Delta})$ and width $(\sigma_{n_{\Delta}})$ increase monotonically with centrality. Within each centrality bin this dependence can be approximated with linear functions given by
\bea
\bar{n}_{\Delta}(N_{\rm ch}) & = & \bar{n}_{\Delta}^0 + \delta\bar{n}_{\Delta}
(N_{\rm ch} - \bar{N}_{\rm ch})
\label{Eq89} \\
\sigma_{n_{\Delta}}(N_{\rm ch}) & = & \sigma_{n_{\Delta}}^0 + \delta\sigma_{n_{\Delta}} (N_{\rm ch} - \bar{N}_{\rm ch}),
\label{Eq90}
\eea
where the centrality trends of the data are well described with the parameters listed in Table~\ref{TableI}. The above power-law and Gaussian distributions were sampled to obtain the event-wise number of positive and negative charged particles within the acceptance.

Nonidentified charged-particle $p_t$ spectra, $d^2N_{\rm ch}/2\pi p_t dp_t d\eta$, were reported by the STAR~\cite{STARspectra} and PHENIX~\cite{PHENIXspectra} Collaborations for Au + Au minimum-bias collisions at 200~GeV. These data can be accurately described in the range 0.15~GeV/$c$ $\leq p_t \leq$ 4~GeV/$c$ by a Levy distribution~\cite{Ayamtmt,Levy} given by
\bea
\frac{d^2N_{\rm ch}}{2\pi p_t dp_t d\eta} & = & \frac{A}{[1+\beta(m_t-m_0)/q]^q} ,
\label{Eq91}
\eea
where $A$, $\beta \equiv 1/T$ and $q$ are fitting parameters, $m_t = \sqrt{p_t^2 + m_0^2}$ and $m_0$ is assumed to be the pion mass. Inverse exponent $q^{-1}$ can be interpreted as the relative variance $\sigma_{\beta}^2 / \bar{\beta}^2$ of an event-wise fluctuating slope parameter $\beta$ of a $p_t$ distribution $\exp{(-\beta m_t)}$~\cite{Ayamtmt}. Fit parameters for the STAR and PHENIX spectra data were interpolated to the centralities defined in Ref.~\cite{axialCI} which were used here. The $p_t$-integrated yields were then normalized to the efficiency corrected $dN_{\rm ch}/d\eta$ for each centrality given in Ref.~\cite{axialCI}. The parameters are listed in Table~\ref{TableII}. The $p_t$ distribution can be expressed as a $y_t$ distribution using $dp_t/dy_t \approx m_t$ near $\eta = 0$ where
\bea
\frac{d^2N_{\rm ch}}{dy_td\eta} & = & 2\pi p_t \frac{dp_t}{dy_t}
\frac{d^2N_{\rm ch}}{2\pi p_t dp_t d\eta}.
\label{Eq92}
\eea
The resulting parent distributions were sampled $n^{\pm}$ times to determine the transverse rapidities for the particles in each simulated event.

\begin{table}[htb]
\caption{Centrality bins, multiplicity ranges, and multiplicity dependent single-particle distribution parameters for the Monte Carlo simulations discussed in the text.}
\label{TableI}
\begin{tabular}{cccc}
\hline \hline
    & \multicolumn{3}{c}{Centrality} \\
\hline
Parameter  & 84-93\%   &  55-64\%   &  9-18\%   \\
\hline
$[N_{\rm ch,min}$,$N_{\rm ch,max}]$ & [2,14]  &  [66,117]  &  [644,910]  \\
$\bar{N}_{\rm ch}$ &  6.6  & 89.7  & 771 \\
$\bar{n}^0_{\Delta}$    &  0.27  & 1.56  & 6.48 \\
$\delta \bar{n}_{\Delta}$ & 0.058  & 0.015  & 0.0038 \\
$\sigma^0_{n_{\Delta}}$ & 1.20    & 6.83   & 21.5   \\
$\delta\sigma_{n_{\Delta}}$ & 0.26 & 0.035 & 0.010 \\
$T_0$ (GeV)             & 0.1540  & 0.1828  & 0.2184 \\
$T_1$ (GeV)             & 0.00075 & 0.000174 & 0.0000224 \\
$q_0$                   & 10.425  & 11.858  & 16.120 \\
$q_1$                   & 0.033  & 0.0124  & 0.00393 \\
\hline \hline
\end{tabular}
\end{table}

\begin{table*}[t]
\caption{Efficiency corrected multiplicity and estimated number of participant nucleons~\cite{axialCI}, single-particle $p_t$ Levy distribution parameters, and 2D Levy distribution parameters for eleven centrality bins for the Monte Carlo simulations discussed in the text.}
\label{TableII}
\begin{tabular}{cccccccccc}
\hline \hline
Centrality(\%) & $\frac{dN_{\rm ch}}{d\eta}$ & $N_{\rm part}$ & $A$~(GeV/$c$)$^{-2}$ &
$T$~(GeV) & $q$ & $\Delta(1/q)^{\rm LS}_{\Sigma}$ & $\Delta(1/q)^{\rm LS}_{\Delta}$ &
$\Delta(1/q)^{\rm US}_{\Sigma}$ & $\Delta(1/q)^{\rm US}_{\Delta}$ \\
\hline
 84-93 &   5.2 &   4.6 &   14.70 & 0.1540 & 10.425 & 0.0058 & -0.0218 & 0.00120 & -0.0236 \\
 74-84 &  13.9 &  10.5 &   35.65 & 0.1647 & 10.822 & 0.0042 & -0.0143 & 0.00129 & -0.0165 \\
 64-74 &  28.8 &  20.5 &   68.33 & 0.1740 & 11.290 & 0.0024 & -0.0087 & 0.00115 & -0.0100 \\
 55-64 &  52.8 &  36.0 &  117.0  & 0.1828 & 11.858 & 0.0016 & -0.0060 & 0.00096 & -0.0068 \\
 46-55 &  89   &  58.1 &  185.3  & 0.1914 & 12.560 & 0.0009 & -0.0047 & 0.00091 & -0.0048 \\
 38-46 & 139   &  86.4 &  275.1  & 0.1989 & 13.321 & 0.0009 & -0.0035 & 0.00062 & -0.0037 \\
 28-38 & 209   & 124.6 &  395.5  & 0.2059 & 14.173 & 0.0007 & -0.0028 & 0.00059 & -0.0028 \\
 18-28 & 307   & 176.8 &  558.4  & 0.2124 & 15.117 & 0.0006 & -0.0022 & 0.00048 & -0.0019 \\
  9-18 & 440   & 244.4 &  772.6  & 0.2184 & 16.120 & 0.0005 & -0.0015 & 0.00040 &
 -0.0018 \\
  5-9  & 564   & 304.1 &  968.0  & 0.2224 & 16.872 & 0.0004 & -0.0013 & 0.00035 & -0.0014 \\
  0-5  & 671   & 350.3 & 1129.7  & 0.2258 & 17.547 & 0.0004 & -0.0012 & 0.00032 & -0.0013 \\
\hline \hline
\end{tabular}
\end{table*}

Systematic bias due to variations in the shape of the single-particle $p_t$ spectrum was simulated by allowing parameters $T$ and $q$ in Eq.~(\ref{Eq91}) to vary within each centrality bin. A linear dependence of $T$ and $q$ with respect to $N_{\rm ch}$ within each centrality bin was assumed based on the trends in Table~\ref{TableII}. Systematic variations, if any, in the separate shapes of the positive and negative charged-particle $p_t$ distributions with respect to charge difference, $n_{\Delta}$, have not been reported. Possible linear (antisymmetric on $n_{\Delta}$) and quadratic (symmetric) dependences were therefore assumed in this study. Within each centrality bin $T$ and $q$ for positive and negative charged particles were described by the functions
\bea
T^{\pm}(N_{\rm ch},n_{\Delta}) & = & T_0 + T_1(N_{\rm ch} - \bar{N}_{\rm ch})
\nonumber \\
 & + &  T_2^{\pm} n_{\Delta} + T_3^{\pm} n_{\Delta}^2/\sigma_{n_{\Delta}}
\label{Eq93} \\
q^{\pm}(N_{\rm ch},n_{\Delta}) & = & q_0 + q_1(N_{\rm ch} - \bar{N}_{\rm ch})
\nonumber \\
 & + &  q_2^{\pm} n_{\Delta} + q_3^{\pm} n_{\Delta}^2/\sigma_{n_{\Delta}}
\label{Eq94}
\eea
where four of the parameter values are listed in Table~\ref{TableI}. Parameters $T_0$, $T_1$, $q_0$ and $q_1$, determined by fitting the centrality-dependent trends in Table~\ref{TableII}, were assumed to be the same for positive and negative charged particles. In lieu of further measurements, parameters $|T_2^{\pm}|$ and $|T_3^{\pm}|$ were assumed equal to the magnitude of $T_1$, i.e. the same variation with particle number. Similarly, $|q_2^{\pm}|$ and $|q_3^{\pm}|$ were set equal to the magnitude of $q_1$. Linear and quadratic variations with $n_{\Delta}$ were studied separately and the relative algebraic signs for positive and negative charged-particle shape variations were alternated where it was assumed that $T^+_{2,3} = \pm T^-_{2,3} = T_1$ and $q^+_{2,3} = \pm q^-_{2,3} = \pm q_1$. The systematic bias effects resulting from the assumed dependence on charge difference should be taken as an estimate of possible systematic bias, pending future measurements of charge-identified $p_t$ distributions with respect to $n_{\Delta}$.

Preliminary $\Delta\rho/\sqrt{\rho_{\rm ref}}$ transverse rapidity correlations for 200~GeV Au + Au collisions were reported by Oldag~\cite{LizThesis,LizHotQuarks} as a function of centrality. Although these correlations are preliminary and precede the present work they provide a reasonable estimate of the expected magnitudes and centrality dependence of the correlations and can be used to study systematic bias. Parametrizations of these data were used to construct sibling pair weights for the Monte Carlo simulations where the $N_{\rm ch}$-dependence was calculated by interpolating each parameter to the selected value of $N_{\rm ch}$. Simulated correlations computed by averaging over finite centrality bin widths will be compared to the input correlation at the centrality-bin mid-point determined by $\bar{N}_{\rm ch}$.

The correlation data in Refs.~\cite{LizThesis,LizHotQuarks} were defined as
\bea
\frac{\Delta\rho}{\sqrt{\rho_{\rm ref}}} & = & 
\frac{S N^{\rm mix}_{kl}}{\sqrt{N^{\rm soft}_{kl}}}
\left[
\frac{\hat{N}^{\rm sib}_{kl} - \hat{N}^{\rm mix}_{kl}}{\hat{N}^{\rm mix}_{kl}}
\right],
\label{Eq95}
\eea
where $S$ is the prefactor scaling coefficient described below and ``hat'' symbols denote unit-normalized distributions as defined previously. In Eq.~(\ref{Eq95}) the ratio in square brackets is equivalent to the total pair-number normalization discussed in Sec.~\ref{SecIIA}. Mixed-event and soft-particle pair distributions are given by
\bea
N^{\rm mix}_{kl} & = & \frac{d^2N_{\rm ch}}{dy_{t,k}d\eta}
\frac{d^2N_{\rm ch}}{dy_{t,l}d\eta},
\label{Eq96} \\
N^{\rm soft}_{kl} & = & \frac{d^2N_{\rm ch,soft}}{dy_{t,k}d\eta}
\frac{d^2N_{\rm ch,soft}}{dy_{t,l}d\eta},
\label{Eq97}
\eea
using Eqs.~(\ref{Eq91}) and (\ref{Eq92}). The soft-particle production spectrum, defined as that part of the single-particle $p_t$ spectrum which scales with $N_{\rm part}$, can be estimated from the $N_{\rm ch} \rightarrow 0$ limit of the shape of the $p_t$ spectrum as explained in Ref.~\cite{TomSpectrum}. Solving Eq.~(\ref{Eq95}) for the sibling-pair distribution gives
\bea
\hat{N}^{\rm sib}_{kl} & = & \hat{N}^{\rm mix}_{kl} \left\{ 1 +
\frac{\sqrt{N^{\rm soft}_{kl}}}{S N^{\rm mix}_{kl}}
\left[ \frac{\Delta\rho}{\sqrt{\rho_{\rm ref}}} \right]_{\rm model} \right\}
\label{Eq98}
\eea
where $[\Delta\rho/\sqrt{\rho_{\rm ref}}]_{\rm model}$ is the model function used to fit the preliminary correlation data in Refs.~\cite{LizThesis,LizHotQuarks}. That model uses a 2D Levy distribution~\cite{Ayamtmt} given by
\bea
N^{\rm sib}_{kl,{\rm 2DLevy}} & = & (2\pi)^2 p_{t,k} p_{t,l} m_{t,k} m_{t,l}
\left( 1 + \frac{\beta m_{t,\Sigma}}{2q_{\Sigma}} \right)^{-2q_{\Sigma}}
\nonumber \\
 & \times & 
\left[ 1 - \left( \frac{\beta m_{t,\Delta}}{2q_{\Delta} + \beta m_{t,\Sigma}}
 \right)^2 \right]^{-q_{\Delta}}
\label{Eq99}
\eea
where $m_{t,\Sigma} = m_{t,k} + m_{t,l} - 2m_0$, $m_{t,\Delta} = m_{t,k} - m_{t,l}$ and inverse exponents are given by
\bea
\Delta(1/q)_{\Sigma} & = & \frac{1}{q_{\Sigma}} - \frac{1}{q},
\label{Eq100} \\
\Delta(1/q)_{\Delta} & = & \frac{1}{q_{\Delta}} - \frac{1}{q}.
\label{Eq101}
\eea
Differences $\Delta(1/q)_{\Sigma,\Delta}$ were shown in Ref.~\cite{Ayamtmt} to represent the covariance of the 2D, event-wise distribution of slope parameters $(\beta_1,\beta_2)$ for arbitrary particles 1 and 2. Numerical values which fit the preliminary LS and US correlation data in Eq.~(\ref{Eq95}) for away-side particle pairs (relative azimuth $>$ $\pi/2$) are listed in Table~\ref{TableII}. For these cases the prefactor scaling coefficient is $1/\sqrt{4} = 1/2$. Sibling-pair weights computed using Eqs.~(\ref{Eq98}) and (\ref{Eq99}) are obtained from
\bea
N^{\rm sib}_{kl} & = & N^{\rm mix}_{kl} \left[ 1 + \left( \hat{N}^{\rm sib}_{kl,{\rm 2DLevy}}
- \hat{N}^{\rm mix}_{kl} \right) / \hat{N}^{\rm mix}_{kl} \right]
\nonumber \\
\label{Eq102}
\eea
where the weight factor in square brackets is of order unity. Equation~(\ref{Eq102}) was calculated for LS and US pairs. Parameters $\beta$, $q$, $q_{\Sigma}$ and $q_{\Delta}$ in Eqs.~(\ref{Eq91}) and (\ref{Eq99}) for arbitrary $N_{\rm ch}$ were interpolated from the values in Table~\ref{TableII}.

Monte Carlo simulations were carried out for each selected centrality using the following steps: (1) The power-law and Gaussian frequency distributions were sampled to obtain $n^+$, $n^-$ and $N_{\rm ch} = n^+ + n^-$ for each event. (2) The single-particle $y_t$ parent distributions were computed using Eqs.~(\ref{Eq91}) - (\ref{Eq94}) for event-wise values of $N_{\rm ch}$ and $n_{\Delta}$ and were then sampled $n^{\pm}$ times for positive/negative charged-particle multiplicities, where parameter $A$ in Eq.~(\ref{Eq91}) was normalized to the event-wise number of particles. (3) Correlation pair weights were calculated from Eqs.~(\ref{Eq96}), (\ref{Eq99}) and (\ref{Eq102}) using interpolated values at variable $N_{\rm ch}$ or at fixed $\bar{N}_{\rm ch}$ for each sibling pair. (4) Sibling pair and mixed-event pair histograms were accumulated for $(++)$, $(--)$, $(+-)$ and $(-+)$ charged-particle pairs and for each correlation definition in Sec.~\ref{SecII}. (5) Event averages were constructed for LS, US, CI, CD, CI-alternate and CD-alternate for correlation definitions based on $\Delta\sigma^2_{p_t:m}$ and its alternate form (Sec.~\ref{SecIIB}), $\Phi_{p_t}^{(0)}$ (Sec.~\ref{SecIIC}), $\sigma^2_{p_t,{\rm dynamical}}$ (Sec.~\ref{SecIID}), and $F_{p_t}$ (Sec.~\ref{SecIIE}). (6) Prefactors (Sec.~\ref{SecIIG}) were then calculated and applied. The resulting correlations for finite centrality bin-widths were compared with that expected in the absence of bias as discussed in the next section.

\begin{table}[h]
\caption{Monte Carlo model parameter settings for the eleven types of simulation runs in this paper. The entry $\Delta N_{\rm ch} > 0$ means that the finite bin widths in Table~\ref{TableI} were used. For $n_{\Delta}$, the notation ``vary'' means that non-zero values of parameters $\bar{n}^0_{\Delta}$, $\delta\bar{n}_{\Delta}$, $\sigma^0_{n_{\Delta}}$ and $\delta\sigma_{n_{\Delta}}$ in Table~\ref{TableI} were used. $T_1$ and $q_1 \neq 0$ refer to the values in Table~\ref{TableI}. Labels ``same,'' ``diff'' and ``mix'' mean that $T^+_2 = T^-_2 = T_1$ and $q^+_2 = q^-_2 = q_1$, $T^+_2 = -T^-_2 = T_1$ and $q^+_2 = -q^-_2 = q_1$, and $T^+_2 = -T^-_2 = T_1$ while $q^+_2 = -q^-_2 = -q_1$, respectively. Similar values apply when $T^{\pm}_3$ and $q^{\pm}_3$ are non-zero. Correlation weights are ``fixed'' when they are calculated for $N_{\rm ch} = \bar{N}_{\rm ch}$ and are ``varied'' when calculated as a function of $N_{\rm ch}$.}
\label{TableIII}
\begin{tabular}{ccccccc}
\hline \hline
Simulation & & & & & & Correlation \\
Run Type  & $\Delta N_{\rm ch}$ & $n_{\Delta}$ & $T_1,q_1$ & $T^{\pm}_2,q^{\pm}_2$ &
$T^{\pm}_3,q^{\pm}_3$ & pair weights \\
\hline
  1  &   0  &   0   &   0,0  &  0,0   &   0,0   &  1.0  \\
  2  & $>0$ & vary  &   0,0  &  0,0   &   0,0   &  1.0  \\
  3  & $>0$ & vary  &$\neq 0$&  0,0   &   0,0   &  1.0  \\
  4  & $>0$ & vary  &$\neq 0$& same   &   0,0   &  1.0  \\
  5  & $>0$ & vary  &$\neq 0$& diff   &   0,0   &  1.0  \\
  6  & $>0$ & vary  &$\neq 0$& mix    &   0,0   &  1.0  \\
  7  & $>0$ & vary  &$\neq 0$&  0,0   &  same   &  1.0  \\
  8  & $>0$ & vary  &$\neq 0$&  0,0   &  diff   &  1.0  \\
  9  & $>0$ & vary  &$\neq 0$&  0,0   &  mix    &  1.0  \\
 10  & $>0$ & vary  &   0,0  &  0,0   &   0,0   &$\neq 1$,fixed \\
 11  & $>0$ & vary  &   0,0  &  0,0   &   0,0   &$\neq 1$,varied \\
\hline \hline
\end{tabular}
\end{table}

\begin{figure}
\includegraphics[keepaspectratio,width=3.5in]{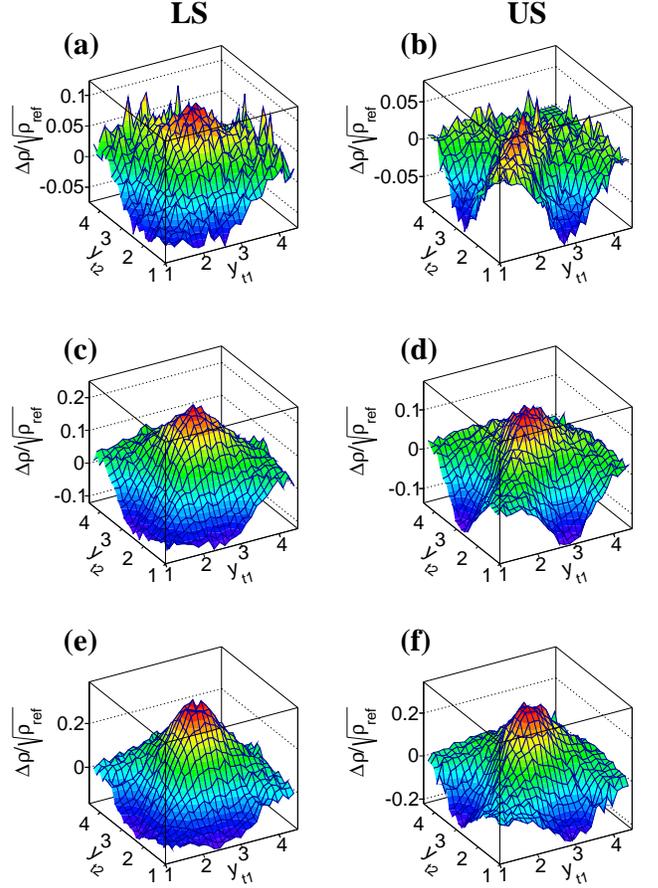}
\caption{\label{Fig1}
Simulated correlations $\Delta\rho/\sqrt{\rho_{\rm ref}}(y_{t1},y_{t2})$ for Au + Au collisions at 200~GeV for LS pairs (panels a, c, e) and US pairs (panels b, d, f) corresponding to peripheral (panels a, b), mid-central (panels c, d) and more-central (panels e, f) collisions. For these calculations finite centrality bin-widths and the correlations from Refs.~~\cite{LizThesis,LizHotQuarks} were assumed as explained in the text.}
\end{figure}

\begin{figure}
\includegraphics[keepaspectratio,width=3.5in]{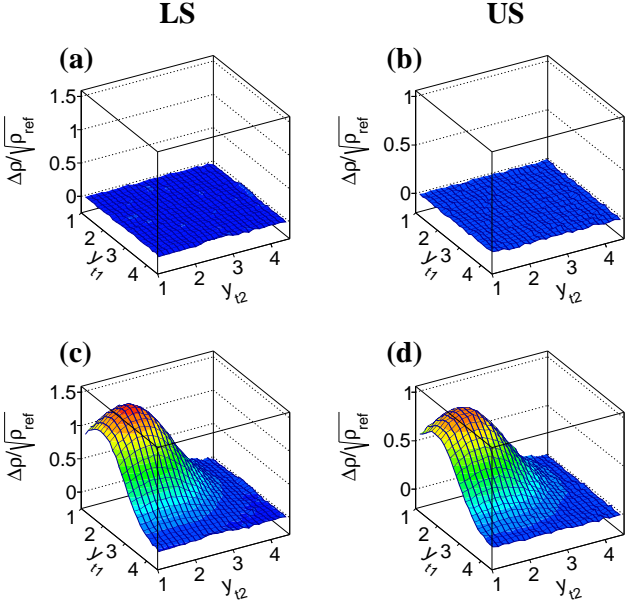}
\caption{\label{Fig2}
Statistical bias effects for event-averaged correlation $\Delta\rho/\sqrt{\rho_{\rm ref}}(y_{t1},y_{t2})$ for 200~GeV mid-central Au + Au collisions. Results assuming fixed $N_{\rm ch}$ and finite multiplicity bin widths with no input correlations are shown in panels (a,b) and (c,d), respectively. Results for LS pairs and US pairs are shown in panels (a,c) and (b,d), respectively.}
\end{figure}

\begin{figure*}
\includegraphics[keepaspectratio,width=6.5in]{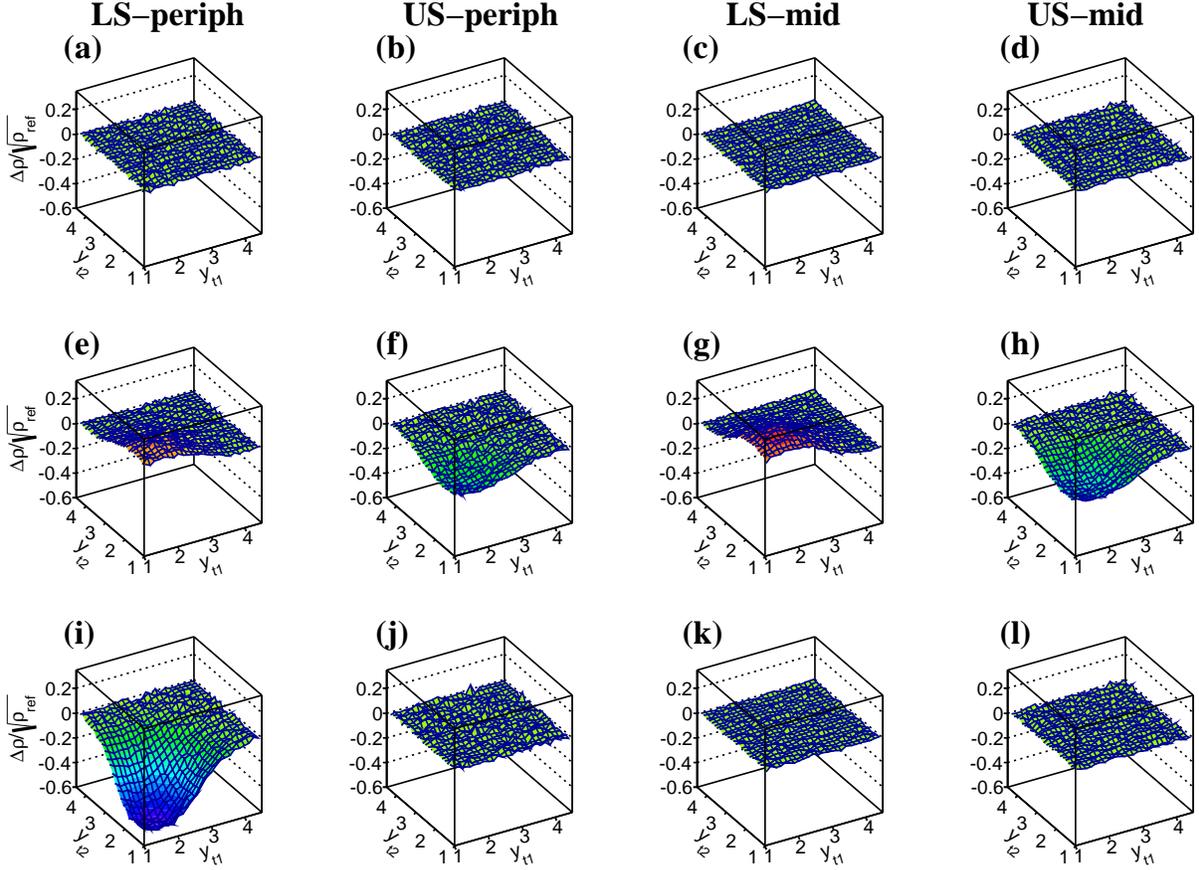}
\caption{\label{Fig3}
Statistical bias effects in $\Delta\rho/\sqrt{\rho_{\rm ref}}(y_{t1},y_{t2})$ for 200~GeV Au + Au collisions due to finite multiplicity bin-widths corresponding to simulation run type~2 in Table~\ref{TableIII}. No input correlations were used. LS pairs in peripheral collisions (panels a, e, i), US peripheral (panels b, f, j), LS mid-central (panels c, g, k), and US mid-central (panels d, h, l) are shown in the columns of panels from left-to-right, respectively. Simulation results based on $\Delta \sigma^2_{p_t:m}$, its alternate form in Eq.~(\ref{Eq26}), and $\sigma^2_{p_t,{\rm dynamical}}$ are shown in successive rows of panels from upper-to-lower.}
\end{figure*}

\begin{figure}
\includegraphics[keepaspectratio,width=3.5in]{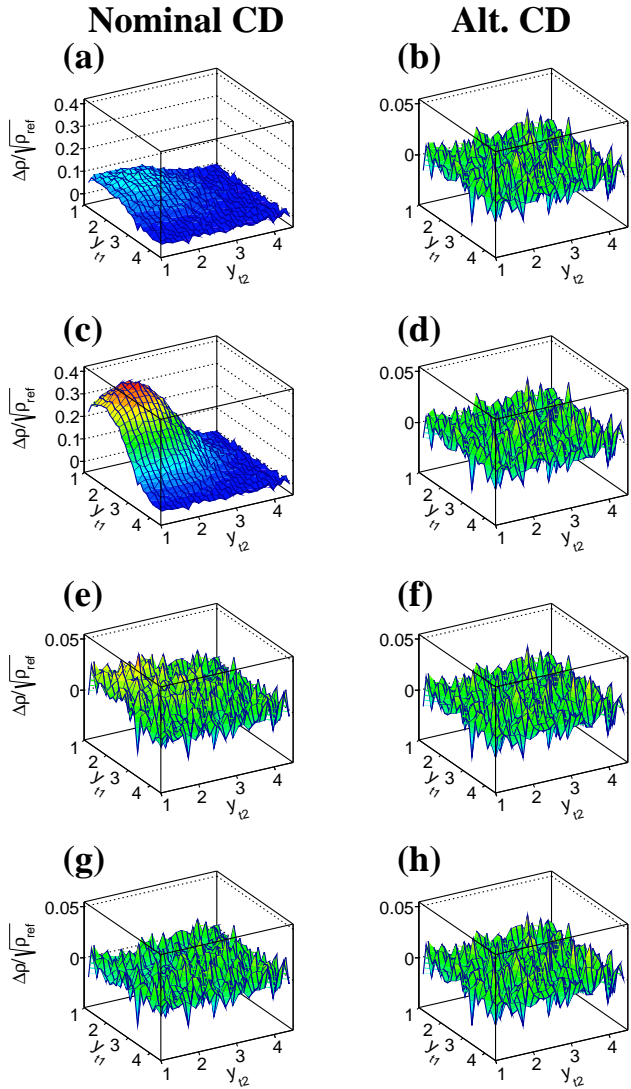}
\caption{\label{Fig4}
Statistical bias effects for the charge-dependent (CD = LS$-$US) $\Delta\rho/\sqrt{\rho_{\rm ref}}(y_{t1},y_{t2})$ quantity for mid-central, 200~GeV Au + Au collisions including finite multiplicity bin-width only with no input correlations. Nominal and alternate CD forms were used for the results shown in the left-hand panels (a, c, e, g) and right-hand panels (b, d, f, h), respectively. Possible bias effects are shown for correlation quantities derived from $\Delta \sigma^2_{p_t:m}$, $\Phi_{p_t}^{(0)}$, $\sigma^2_{p_t,{\rm dynamical}}$, and $F_{p_t}$ in successive rows of panels from upper-to-lower.}
\end{figure}

\begin{figure*}
\includegraphics[keepaspectratio,width=6.5in]{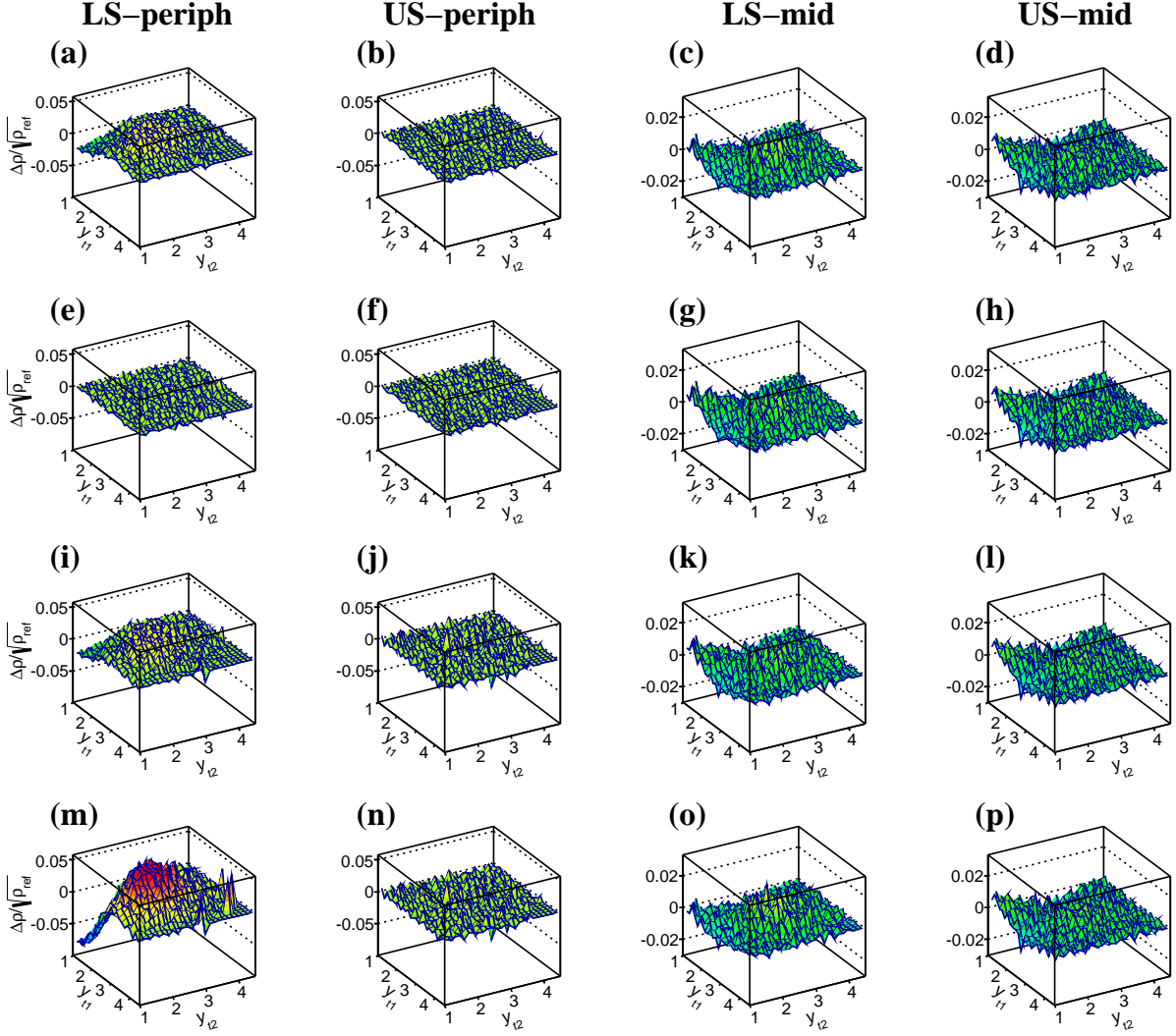}
\caption{\label{Fig5}
Systematic bias effects for 200~GeV Au + Au collisions due to the $N_{\rm ch}$-dependent shape of the single-particle $p_t$ distribution as discussed in the text. LS and US pair correlations for peripheral and mid-central collisions are shown in the columns of panels as in Fig.~\ref{Fig3}. Results are shown for correlations derived from $\Delta \sigma^2_{p_t:m}$ (panels a-d), $\Phi_{p_t}^{(0)}$ (panels e-h), $\sigma^2_{p_t,{\rm dynamical}}$ (panels i-l), and $F_{p_t}$ (panels m-p).}
\end{figure*}

\begin{figure*}
\includegraphics[keepaspectratio,width=6.5in]{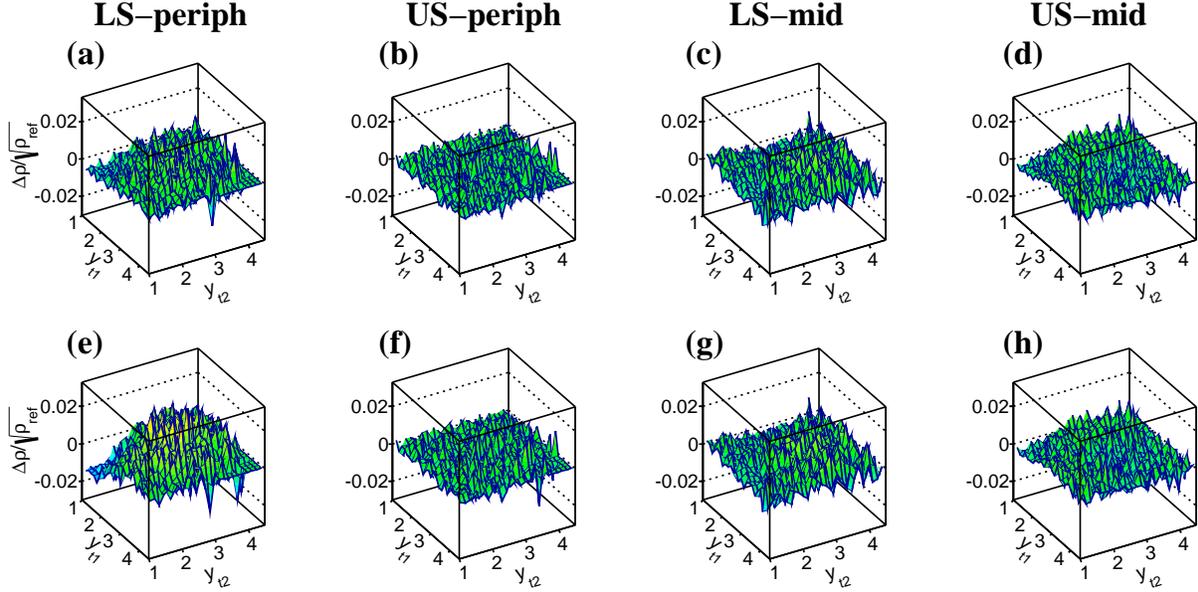}
\caption{\label{Fig6}
Systematic bias effects for 200~GeV Au + Au collisions due to possible $n_{\Delta}$ dependence in the positive and negative single-particle $p_t$ distributions as discussed in the text. LS and US pair correlations for peripheral
 and mid-central collisions are shown in the columns of panels as in Fig.~\ref{Fig3}. Results are shown for correlations derived from $\Delta \sigma^2_{p_t:m}$ (panels a-d) and $F_{p_t}$ (panels e-h).}

\end{figure*}

\begin{figure*}
\includegraphics[keepaspectratio,width=6.5in]{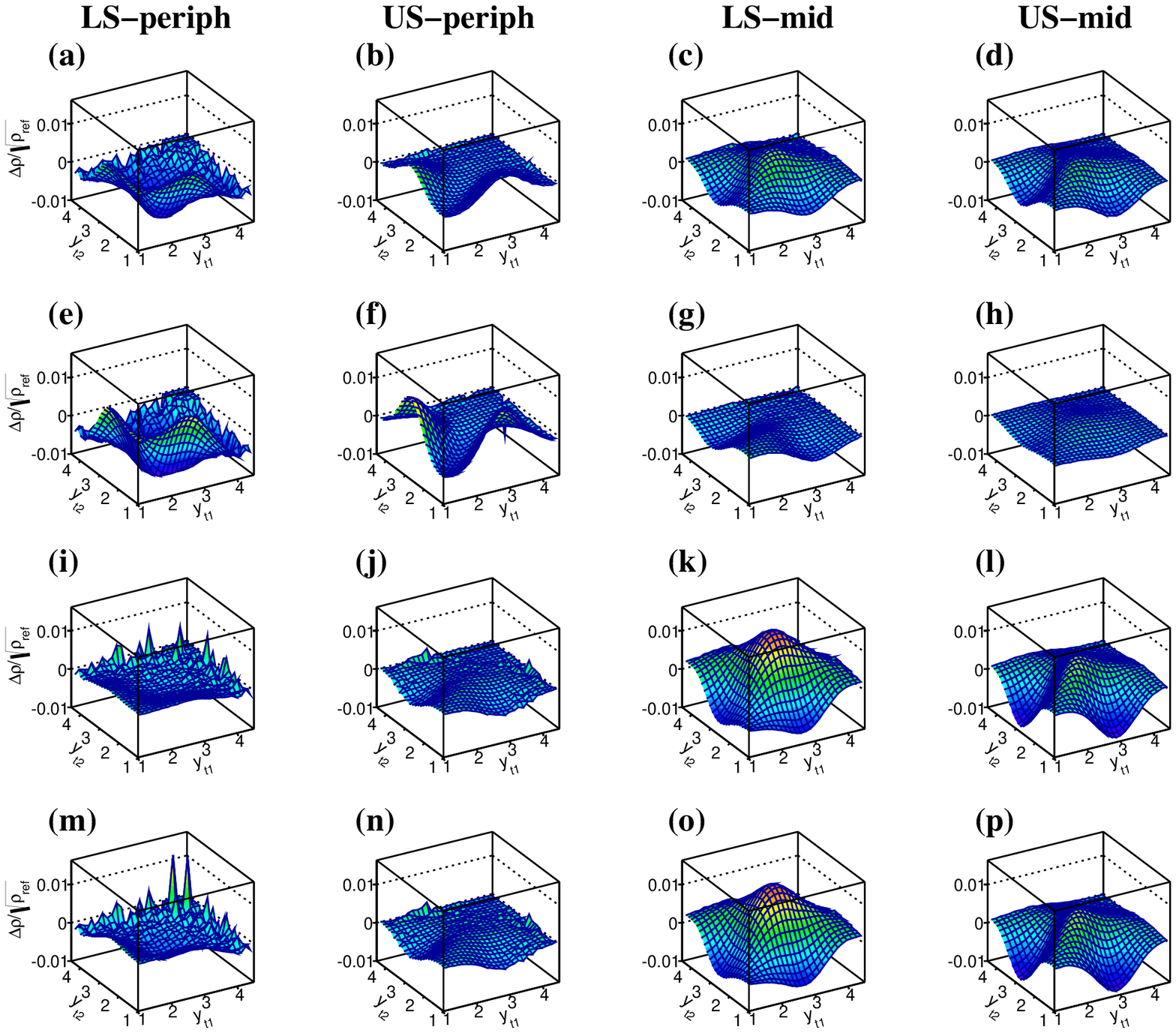}
\caption{\label{Fig7}
Systematic bias effects for 200~GeV Au + Au collisions due to $N_{\rm ch}$-dependence of the correlation shapes as discussed in the text. LS and US pair correlations for peripheral and mid-central collisions are shown in the columns of panels as in Fig.~\ref{Fig3}. Results are shown for correlations derived from $\Delta \sigma^2_{p_t:m}$ (panels a-d), $\Phi_{p_t}^{(0)}$ (panels e-h), $\sigma^2_{p_t,{\rm dynamical}}$ (panels i-l), and $F_{p_t}$ (panels m-p).}
\end{figure*}

\section{Results}
\label{SecV}

For the correlation quantities in Sec.~\ref{SecII} eleven types of simulation calculations were done for the peripheral, mid-central and more-central bins using $10^6$, $10^6$ and $0.5\times10^6$ collision events in each run, respectively. The event sample sizes were sufficient to ensure that statistical and/or systematic bias effects large enough to compromise correlation measurements are clearly visible in the simulations. Typical statistical errors in the simulated $\Delta\rho/\sqrt{\rho_{\rm ref}}$ are of order $\pm0.01$, $\pm0.01$ and $\pm0.015$ for the peripheral, mid-central and more-central collisions, respectively. These errors are about one-tenth of the expected correlation magnitudes~\cite{LizThesis,LizHotQuarks}. The eleven simulation runs progressively included additional bias producing effects. The first assumed fixed $N_{\rm ch}$ with $n_{\Delta}$ = 0 and no correlation weights. Subsequent calculations included finite multiplicity bin widths and non-zero $n_{\Delta}$. Then $N_{\rm ch}$-dependent single-particle $p_t$ spectrum shapes were added, followed by $n_{\Delta}$-dependent $p_t$ spectrum shapes for positive and negative charged particles. Finally, $N_{\rm ch}$-dependent correlation shape variation was included. Explicit parameter settings for the eleven types of calculations are explained in Table~\ref{TableIII}.


Simulated correlations $\Delta\rho/\sqrt{\rho_{\rm ref}}(y_{t1},y_{t2})$ for each centrality and for like-sign and unlike-sign charged-particle pairs are shown in Fig.~\ref{Fig1} corresponding to simulation run type~10 in Table~\ref{TableIII} and assuming the $\Delta \sigma^2_{p_t:m}$ based correlation. Single-particle spectrum shapes and 2D Levy parameters were fixed at their respective $\bar{N}_{\rm ch}$ values in each centrality bin. The essential features of the correlations include: (1) an overall increase in the amplitude with increasingly more-central collisions, (2) the evolution of the shape of the unlike-sign correlations along the diagonal $y_{t1} \approx y_{t2}$ bins, and (3) the prominent peak at $(y_{t1},y_{t2}) \approx (3,3)$. The magnitudes of the bias effects discussed below may be compared to these estimated correlations.

In Fig.~\ref{Fig2} simulations for the mid-central bin assuming the event-averaged correlation form in Eq.~(\ref{Eq10}) are shown for LS and US pairs in upper panels (a) and (b) for fixed multiplicity $\bar{N}_{\rm ch}$ and no input correlation (simulation run type~1). The results are statistically consistent with zero indicating no bias. In the lower panels LS and US results for finite multiplicity bin-width and no input correlations (simulation run type~2) are shown where large structure appears indicating strong statistical bias. In the absence of correlation weights [weight factor in Eq.~(\ref{Eq102}) equals unity] quantity $\Delta\rho/\sqrt{\rho_{\rm ref}}$ should ideally be statistically consistent with zero. Statistically significant non-zero correlations resulting from simulation run types~2 - 9 in Table~\ref{TableIII} indicate the presence of bias. Fig.~\ref{Fig2} shows that the simple, event-averaged correlation induces large, statistical bias for both LS and US pairs for finite width multiplicity bins, as shown in Eq.~(\ref{Eq10}). Applications of this event-averaged correlation require the variance of the selected event-wise multiplicity fluctuations to be much less than $\bar{N}^2$.

Statistical biases due to pair-counting in finite-width multiplicity bins (simulation run type~2) for correlations derived from fluctuation quantities $\Delta \sigma^2_{p_t:m}$ and its alternate charge-identified form in Eq.~(\ref{Eq26}), $\Phi_{p_t}^{(0)}$, $\sigma^2_{p_t,{\rm dynamical}}$, and $F_{p_t}$ were studied for LS and US charged-particle pairs and for the three centrality bins. Results for $\Delta \sigma^2_{p_t:m}$, $\Phi_{p_t}^{(0)}$ and $F_{p_t}$ were statistically consistent with zero (unbiased) as shown for the $\Delta \sigma^2_{p_t:m}$ results in the upper row of panels in Fig.~\ref{Fig3}. Results for the alternate, charge-identified form in Eq.~(\ref{Eq26}) are strongly biased (i.e. bias effects are larger than the expected correlations) as shown in the middle row of panels in this figure. Only the charge-nonidentified results (not shown) from Eq.~(\ref{Eq20}) are statistically unbiased. The explicit treatment of unlike-sign charged-particle pairs as in Eq.~(\ref{Eq23}) is essential for eliminating statistical bias.
Results for the quantity based on $\sigma^2_{p_t,{\rm dynamical}}$ (lower row of panels in Fig.~\ref{Fig3}) are highly biased for the peripheral, LS correlations but are statistically unbiased for the other cases. The large bias effect can be avoided by restricting $N_{\rm ch} > 2$, based on calculations with $N_{\rm ch} \in [3,14]$ (not shown). Additional simulations with $N_{\rm ch} \in [1,14]$ (also not shown) indicate that large statistical bias appears in the LS correlations obtained from $\Delta \sigma^2_{p_t:m}$ and $F_{p_t}$, requiring $N_{\rm ch} > 1$ for these quantities. For high multiplicity events and/or large angular-bin size (scale) these restrictions are of little practical importance. However, for scale-dependent analysis~\cite{ptscale,STARscale} in which the average angular-bin occupancy may be small ($\sim$2), the $N_{\rm ch} >$ 1 or 2 restrictions would distort the event sample used to compute the correlations. Results based on $\Phi_{p_t}^{(0)}$ are statistically unbiased for bin-wise occupancies with $N_{\rm ch} \geq 1$. The $\Phi_{p_t}^{(0)}$ quantity was used in the scale-dependent, mean-$p_t$ fluctuation analysis of STAR data in Refs.~\cite{ptscale,STARscale}.

In Eqs.~(\ref{Eq76}) and (\ref{Eq77}) two forms for the charge-dependent (CD = LS$-$US) correlation are given. The results for mid-central collisions with finite multiplicity bin width (simulation run type~2, no input correlation) are shown in Fig.~\ref{Fig4}. The nominal [Eq.~(\ref{Eq76})] and alternate [Eq.~(\ref{Eq77})] CD results are shown in the left and right columns of panels, respectively. Bias effects for the correlation forms derived from fluctuation quantities $\Delta \sigma^2_{p_t:m}$, $\Phi_{p_t}^{(0)}$, $\sigma^2_{p_t,{\rm dynamical}}$ and $F_{p_t}$ are shown in descending order from upper to lower rows, respectively. For the nominal CD results large bias occurs for the $\Delta \sigma^2_{p_t:m}$ and $\Phi_{p_t}^{(0)}$ forms. A small degree of bias is present at lower $y_t$ for the $\sigma^2_{p_t,{\rm dynamical}}$ form. The $F_{p_t}$ results are consistent with zero (unbiased). For each quantity the alternate CD results shown here are unbiased. Note that any difference in the biases for LS and US pairs will contribute directly to both the nominal and alternate CD correlations. The point of Fig.~\ref{Fig4} is to show that the LS$-$US difference computed using the sibling minus mixed differences given in Eq.~(\ref{Eq77}) minimizes the bias and is therefore the recommended form to use for charge-dependent (LS$-$US) correlations.

No significant differences were found between the simulated correlation results for the nominal and alternate charge-independent (CI=LS+US) correlations in Eqs.~(\ref{Eq74}) and (\ref{Eq75}), respectively. Similar results for these two forms were produced for the respective correlation methods discussed in Sec.~\ref{SecII} as all of the statistical and systematic bias sources were successively added to the simulations. The mathematical differences between the two CI correlation forms are contained in the charge-identified weights, given by 1/4 for the nominal CI and $N_{kl}^{{\rm mix},ab}/\sum_{a^{\prime},b^{\prime}=\pm \pm} N_{kl}^{{\rm mix},a^{\prime} b^{\prime}}$ for the alternate CI. When the positive and negative charged-particle $p_t$ distributions differ in shape, the alternate CI weight factors vary with $(y_{t1},y_{t2})$. The effects of these variations are insignificant in the present examples.

Systematic bias due to $N_{\rm ch}$-dependence of the single-particle $p_t$ spectrum shape is shown in Fig.~\ref{Fig5} for LS and US charged-particle pairs for peripheral and mid-central collisions. The panels show differences for simulation run type~3 for $\Delta\rho/\sqrt{\rho_{\rm ref}}$ minus that for run type~2 (no input correlations). Systematic bias effects for $\Delta\rho/\sqrt{\rho_{\rm ref}}(y_{t1},y_{t2})$ derived from $\Delta \sigma^2_{p_t:m}$, $\Phi_{p_t}^{(0)}$, $\sigma^2_{p_t,{\rm dynamical}}$ and $F_{p_t}$ are shown in successive rows of panels from the upper-most row to the bottom row, respectively. Modest bias effects (increases) are seen at lower $y_t$ in all cases for the mid-central collisions. Larger bias is seen for LS pairs in peripheral collisions for the $\Delta \sigma^2_{p_t:m}$ and $\sigma^2_{p_t,{\rm dynamical}}$ based forms. A larger systematic bias appears in LS pair, peripheral collisions for the $F_{p_t}$ based correlation. This bias is as large as the expected correlation signal (see Fig.~\ref{Fig1}). No significant systematic bias effects were found for US pairs in peripheral collisions. 

Systematic bias effects due to assumed variations in the $p_t$ distribution shapes for positive and negative charged particles as a function of $n_{\Delta} = n^+ - n^-$ were estimated using the different sets of model parameters described in Sec.~\ref{SecIV}. The largest estimated bias effects resulted from assuming non-zero values for either parameters $T^{\pm}_2$ and $q^{\pm}_2$, or $T^{\pm}_3$ and $q^{\pm}_3$, in the ``mix'' configuration corresponding to simulation run types~6 or 9 in Table~\ref{TableIII}. Results for non-zero $T^{\pm}_2$ and $q^{\pm}_2$ (``mix''), where $\Delta\rho/\sqrt{\rho_{\rm ref}}$ for run type~3 was subtracted from that for run type~6, are shown in Fig.~\ref{Fig6}. The columns of panels show results for LS and US charged-particle pairs for peripheral and mid-central collisions as in Fig.~\ref{Fig3}. Bias results for correlations derived from $\Delta \sigma^2_{p_t:m}$, $\Phi_{p_t}^{(0)}$ and $\sigma^2_{p_t,{\rm dynamical}}$ are similar with small bias effects at lower $y_t$ as shown for $\Delta \sigma^2_{p_t:m}$ in the upper row of panels. Bias effects for the correlation quantity based on $F_{p_t}$ (lower row of panels) are similar except for LS, peripheral which are larger.

In Fig.~\ref{Fig7} the systematic bias due to $N_{\rm ch}$-dependence in the assumed correlation shape [see the pair weight-factor in Eq.~(\ref{Eq102})] relative to the correlation at the mid-point of the multiplicity bin at $\bar{N}_{\rm ch}$, is shown for LS and US charged-particle pairs for peripheral and mid-central collisions. Specifically, the results shown correspond to $\Delta\rho/\sqrt{\rho_{\rm ref}}$ computed in simulation run type~10 subtracted from that for run type~11 (see Table~\ref{TableIII}). The bias effects vary in shape and overall amplitude for LS versus US charged-particle pairs, for peripheral versus mid-central collisions, and for each correlation measure. In general, these systematic bias effects are negligible relative to the expected correlation magnitudes.

Finally, systematic bias effects for the more-central collisions listed in Tables~\ref{TableI} and \ref{TableII} were found to be approximately twice as large as those shown here for mid-centrality. In realistic correlation analyses~\cite{MikeThesis,LizThesis,axialCI} broad multiplicity bin-widths such as the more-central bin considered here with $N_{\rm ch} \in [644, 910]$ must be sub-divided in order that event-mixing for the reference pair densities does not produce artifacts in the correlation structure. Typically, the maximum allowed range for $N_{\rm ch}$ is 50 which would reduce systematic bias effects to levels no greater than those shown in Figs.~\ref{Fig5} - \ref{Fig7}. Statistical biases for the more-central collisions for correlations derived from  $\Delta \sigma^2_{p_t:m}$, $\Phi_{p_t}^{(0)}$, $\sigma^2_{p_t,{\rm dynamical}}$ and $F_{p_t}$ are negligible, even for the full bin-width.

\begin{table}[h]
\caption{Range of applicability of LS and US charged particle pair correlations based on four mean-$p_t$ fluctuation quantities in terms of the allowed range of multiplicity ($N_{\rm ch} = n^+ + n^-$) within an arbitrary, angular bin.}
\label{TableIV}
\begin{tabular}{ccc}
\hline \hline
Fluctuation & &  \\
Quantity  & LS  & US \\
\hline
  $\Delta \sigma^2_{p_t:m}$  &  $N_{\rm ch} > 1$  &  $N_{\rm ch} \geq 1$ \\
  $\Phi_{p_t}^{(0)}$         &  $N_{\rm ch} \geq 1$ &  $N_{\rm ch} \geq 1$ \\
  $\sigma^2_{p_t,{\rm dynamical}}$ &  $N_{\rm ch} > 2$  &  $N_{\rm ch} \geq 1$ \\
  $F_{p_t}$                  &  $N_{\rm ch} > 1$  &  $N_{\rm ch} \geq 1$ \\
\hline \hline
\end{tabular}
\end{table}

The ranges of applicability of four of the LS and US correlation quantities derived here are summarized in Table~\ref{TableIV}. Multiplicity, $N_{\rm ch} = n^+ + n^-$, refers to the event-wise, charged-particle occupancy in arbitrary $(\eta,\phi)$ bins. For the present analysis the $(\eta,\phi)$ bin size was assumed to be equal to the full acceptance of the STAR TPC tracking detector. Requiring $N_{\rm ch} >$ 1 (quantities $\Delta \sigma^2_{p_t:m}$ and $F_{p_t}$ for LS) or $N_{\rm ch} >$ 2 (quantity $\sigma^2_{p_t,{\rm dynamical}}$ for LS) only affects the most-peripheral collision centrality bin in these cases. In general, however, the $(\eta,\phi)$ bin size can be less than the acceptance as explained in Refs.~\cite{ptscale,STARscale}. The $(\eta,\phi)$ bin-wise multiplicity requirement, $N_{\rm ch} >$ 1 or 2, for these three correlation quantities could affect more central collisions and, depending on the collision system and energy and the angular bin-scale, could cause the correlation analyses for these three quantities for LS particle pairs to be unreliable for much of the centrality range. Correlations derived from $\Phi_{p_t}^{(0)}$ do not suffer from the above restrictions on $N_{\rm ch}$.
Statistical bias in the simple, event-averaging correlation (Sec.~\ref{SecIIA}) and in the recent NA49 fluctuation quantity (Sec.~\ref{SecIIF}) can only be eliminated by restricting the event acceptance to nearly zero multiplicity width such that $\sigma^2_N$ is negligible. Systematic biases of the types discussed here can be reduced by restricting the centrality range of the events. 

\section{Summary and Conclusions}
\label{SecVI}

Two-particle correlations on transverse momentum $(p_{t1},p_{t2})$, or transverse rapidity $(y_{t1},y_{t2})$, contain additional, independent information beyond that accessible with angular correlation measurements. These correlations therefore play an important role in efforts to understand the dynamics involved in relativistic heavy-ion collisions. It is essential that transverse momentum correlation measurements, which can be vulnerable to bias effects in the form of distorted shapes and structures, are as free of statistical and systematic biases as possible.

Several correlation quantities were studied, most of which were derived from non-statistical mean-$p_t$ fluctuation measurement quantities in the literature. Bias effects were studied both analytically and numerically via Monte Carlo simulations for Au + Au collisions at $\sqrt{s_{NN}}$ = 200~GeV. For the simulations, event multiplicity distributions, $p_t$-spectrum parameters, and estimated correlation distributions were taken from measurements reported in the literature.

The simple correlation definition based on pair-number normalization, Eq.~(\ref{Eq1}), includes unknown distortions while that based on event-averaging, Eq.~(\ref{Eq10}), includes large statistical bias when event collections are used which have a finite range of multiplicities. Five distinctly different correlation quantities were then studied which were derived from mean-$p_t$ fluctuation quantities $\Delta \sigma^2_{p_t:m}$, $\Phi_{p_t}^{(0)}$, $\sigma^2_{p_t,{\rm dynamical}}$, $F_{p_t}$, and $\Delta [P_T,N]$, $\Sigma [P_T,N]$ in order to ascertain the statistical and systematic biases associated with each.

A simplified charge-identified correlation form based on the charge-nonidentified $\Delta \sigma^2_{p_t:m}$ fluctuation quantity was found to have large statistical bias which exceeded the expected magnitude of the correlation signal. A charge-identified correlation form derived from an explicit charge-identified $\Delta \sigma^2_{p_t:m}$ definition did not contain significant statistical bias. Explicit charge identification was therefore included in all of the other mean-$p_t$ fluctuation quantities considered here.

Statistical bias for like-sign pairs can be problematic (bias effects as large or larger than the expected correlations) for the $\Delta \sigma^2_{p_t:m}$, $\sigma^2_{p_t,{\rm dynamical}}$ and $F_{p_t}$ based correlations for peripheral collisions or for scale-dependent analyses at any centrality where the event-wise bin occupancies can be as low as 1, 2, and 1, respectively. Statistical bias is not an issue for the like-sign and unlike-sign charged-pair correlations based on $\Phi_{p_t}^{(0)}$; the same is true for unlike-sign correlations for the above three correlation quantities. The applicable ranges in $(\eta,\phi)$ bin-wise multiplicities for the LS and US correlations derived from mean-$p_t$ fluctuation quantities $\Delta \sigma^2_{p_t:m}$, $\Phi_{p_t}^{(0)}$, $\sigma^2_{p_t,{\rm dynamical}}$, and $F_{p_t}$ are listed in Table~\ref{TableIV}. Statistical bias in the correlations derived from $\Delta [P_T,N]$ and $\Sigma [P_T,N]$ can only be controlled by severely limiting the bin-wise multiplicity range such that $\sigma^2_N$ is negligible. 

Systematic bias due to multiplicity, or centrality, dependence in the single-particle $p_t$ spectrum shape is more evident at lower $(y_{t1},y_{t2})$ and, in the present simulations, is largest for the like-sign, peripheral collision correlations based on quantity $F_{p_t}$. Systematic bias caused by the overall multiplicity dependence of the correlation shape varies with correlation model, charge pair combination, location in $(y_{t1},y_{t2})$, and centrality. For each case studied here this bias is one-tenth or less of the expected correlation magnitudes. Systematic bias magnitudes are proportional to centrality bin-width and therefore can be reduced by limiting the accepted centrality range in the data analysis. Reducing the centrality bin-width for statistically unbiased quantities until stable correlations are achieved is a straightforward way to minimize this type of systematic bias.

Charge-dependent (CD = LS$-$US) correlations were also studied for each correlation quantity. The like-sign minus unlike-sign sibling pair difference form in Eq.~(\ref{Eq76}) produces spurious results in most cases. The correlated-pair difference form in Eq.~(\ref{Eq77}), $[(\Delta N_{kl}^{++} + \Delta N_{kl}^{--}) - (\Delta N_{kl}^{+-} + \Delta N_{kl}^{-+})]$, does not produce any additional bias beyond that already present in the LS and US charged-particle pair correlations.

In conclusion, two-particle correlations on transverse momentum derived from mean-$p_t$ fluctuation quantities $\Delta \sigma^2_{p_t:m}$, $\Phi_{p_t}^{(0)}$, $\sigma^2_{p_t,{\rm dynamical}}$, and $F_{p_t}$ reproduce intrinsic correlation structure at the mid-point of the centrality bin with reasonable fidelity, if the event-wise range of multiplicities ($N_{\rm ch}$) in the $(\eta,\phi)$ angular bin are appropriately restricted as in Table~\ref{TableIV}, and if the collision centrality bin-width is sufficiently narrow. Multiplicity restrictions are an issue for low event-multiplicities and for $(y_{t1},y_{t2})$ correlation analysis at any collision centrality when mean-$p_t$ fluctuations are measured in small $(\eta,\phi)$ bins. For applications to other collision systems, energies, detector acceptances or angular bin scales than that studied here, the impact of the multiplicity restrictions in Table~\ref{TableIV} should be evaluated and the stability of the correlations with respect to centrality bin-width investigated. The analytical analysis and Monte Carlo simulations presented here can be readily extended to other such applications in order to facilitate the reduction of bias in two-particle correlations on transverse momentum.

\vspace{0.2in}
{\bf Acknowledgements}
\vspace{0.1in}

The authors would like to thank Professor Tom Trainor of the Univ. of Washington for many informative discussions relevant to this work. This research was supported in part by the Office of Science of the U. S. Department of Energy under Grants No. DE-FG02-94ER40845 and No. DE-SC0013391.


\end{document}